\documentclass[12pt]{article}
\usepackage{eqsection,latexsym,epsf}
\usepackage{epsfig,amsfonts,cite}

\begin{document}

\newcommand{\<}{\langle}
\renewcommand{\>}{\rangle}
\def\t{t_{\beta}}
\def\l{\lambda}
\def\lh{\widehat{\lambda}}
\def\ph{\widehat{\pi}}
\def\ux{\underline{x}}
\def\uz{\underline{z}}
\def\u0{\underline{0}}
\def\us{\underline{\sigma}}
\def\wus{\widehat{\us}}
\def\ws{\widehat{\sigma}}
\def\dh{\widehat{\delta}}
\def\d{\delta}
\def\ps{\vec{\sigma}}
\def\pt{\vec{\tau}}
\def\ut{\underline{\tau}}
\def\atanh{{\rm arctanh}}

\title{The glassy phase of Gallager codes}

\author{       
  { Andrea Montanari}              \\
  {\small\it Laboratoire de Physique Th\'{e}orique de l'Ecole Normale
  Sup\'{e}rieure\footnote {UMR 8549, Unit{\'e}   Mixte de Recherche du 
Centre National de la Recherche Scientifique et de 
l' Ecole Normale Sup{\'e}rieure. } }
  \\[-0.2cm]
  {\small\it 24, rue Lhomond, 75231 Paris CEDEX 05, FRANCE}        \\[-0.2cm]
  {\small Internet: {\tt Andrea.Montanari@lpt.ens.fr}}
          \\[-0.1cm]
  {\protect\makebox[5in]{\quad}}  
  \\
}


\maketitle
\thispagestyle{empty}   

\abstract{Gallager codes are the best error-correcting codes to-date.
In this paper we study them by using the tools
of statistical mechanics. The corresponding statistical mechanics model is a 
spin model on a sparse random graph. The model can be solved by elementary
methods (i.e. without replicas) in a large connectivity limit.
For low enough temperatures it presents a completely frozen glassy
phase ($q_{EA}=1$).
The same scenario is shown to hold for finite connectivities. 
In this case we adopt the replica approach and exhibit a 
one-step replica symmetry breaking order parameter.
We argue that our ansatz yields the exact solution of the model.
This allows us to determine the whole phase diagram
and to understand the performances of Gallager codes.}

\vspace{5.cm}
\begin{flushleft} 
LPTENS 01/19    
\end{flushleft}

\clearpage

\section{Introduction}

Information theory \cite{Cover,Viterbi} deals with the problem of
reliable communication
through an imperfect (noisy) communication channel.
This can be done by properly encoding the information message in
such a way to increase its redundancy. If a transmission error occurs
due to the noise, the correct message can be restored by exploiting
this redundancy.

The price to pay for error-correction to be possible is to 
increase the length of the transmitted message, i.e. to decrease the 
information rate through the channel.
In 1948 C.~E.~Shannon \cite{Shannon} computed the maximal achievable rate 
at which information can be transmitted through a given communication channel
(the so-called {\it capacity} of the channel).
Since then a lot of work has been spent for constructing
practical error-correcting codes that could realize Shannon 
prediction, i.e. that could saturate the channel capacity. 

In the past few years it has become progressively clear that such 
an objective is not unreachable.
It has become possible to construct
error-correcting codes which remain effective extremely near to the
Shannon capacity \cite{Chung}.
The reasons of this revolution have been the invention of ``turbo
codes'' \cite{PrimoBerrou}
and the re-invention of ``low-density parity check codes''
(LDPCC) \cite{MacKay}.
The last ones \cite{Gallager}
were proposed for the first time by R.~Gallager in 1962,
but were soon forgotten afterwards, probably because of the lack of
computational resources at that time.

As it has been shown by N.~Sourlas \cite{Sourlas1,Sourlas2,Sourlas4}, 
error-correcting codes can be
mapped onto disordered spin models. This mapping 
allows to employ statistical mechanics techniques to investigate the 
behavior of the former.
Both turbo codes \cite{Turbo1,Turbo2} 
and LDPCC \cite{KanterSaad_PRL,KanterSaad_Cascading,Saad_MN_PRL,
KanterSaad_FS,SaadRegular,SaadCactus,SaadTighter} 
have been already studied using this
approach. However all previous studies were restricted 
to particular regions of the phase diagram.
The principal technical reason was the difficulty of implementing
replica symmetry breaking in finite connectivity systems.
 
In this work we focus on regular Gallager codes (a particular family 
of LDPCC), and we address the fundamental problem of determining the
corresponding phase diagram. 
There are two type of motivations for such a task to be undertaken.
First, the spin model corresponding to Gallager codes is a disordered
spin model on a diluted graph. The study of such systems has
greatly improved our understanding of glassy systems over the last few
years.
Second, it is of great practical importance to have a complete
quantitative picture of the behavior of Gallager codes.
For instance, the existence of a glassy phase can have important
effects on the decoding algorithms, and the knowledge of the phase
diagram can be used to improve them. 

The model is presented Sec. \ref{ModelSection}. 
In Sec. \ref{NishimoriSection} we prove some exact properties which
hold at inverse temperature $\beta = 1$. The line $\beta = 1$
can be regarded as the Nishimori line \cite{Nishimori1} 
of the phase diagram. 
In Sec. \ref{RandomCodewordSection} we solve the model in the 
large connectivity
limit. We show that it becomes identical to a simplified model which
we call the {\it random codeword model} (RCM).
The RCM is shown to have a freezing phase transition
analogous to the one of the random energy model (REM) \cite{DerridaREM}.
In Sec. \ref{ReplicaSection} we adopt the replica approach 
\cite{SpinGlass}
and prove that the same scenario applies for finite connectivities.
In particular we construct a replica symmetry breaking solution of the
saddle point equations.
The proposed solution is much simpler than the generic 
one-step replica symmetry breaking solution. 
Rather than being parametrized by a functional over a probability
space \cite{MonassonRSB}, it depends simply upon the probability 
distribution of a local field. 
Such a probability distribution can be easily computed numerically. It 
can be also obtained from a large connectivity expansion,
see Sec. \ref{ExpansionSection}. In Sec. \ref{FiniteSizeSection}
we compute the finite-size corrections of the free energy 
for the RCM, and compare the result with exact enumerations.
Finally in Sec. \ref{DiscussionSection} we discuss the validity of our
replica symmetry breaking ansatz.
%
%
\section{The model}
\label{ModelSection}

Let us suppose we want to transmit an information message consisting of 
$L$ bits. There are $2^L$ such messages. Each of them is encoded in
a string of $N>L$ bits ({\it codewords}).

This motivates the following model. 
There are $2^L$ possible configurations of the system (the codewords), 
each one corresponding to a distinct sequence of $N>L$ bits. 
We shall denote the
codewords as $\ux^{(\alpha)} = (x^{(\alpha)}_1,\dots,x^{(\alpha)}_N)$,
with $\alpha = 1,\dots,2^L$.
The set of codewords ${\cal C}$ is a linear space. This means
that $\u0 \equiv (0,\dots,0)\in {\cal C}$, and that, if 
$\ux^{(\alpha)},\ux^{(\beta)}\in {\cal C}$, then 
$\ux^{(\alpha)}+\ux^{(\beta)} \in {\cal C}$ (where the sum has to be 
carried modulo $2$).

Like any linear space, the set of codewords ${\cal C}$ can be specified
as the kernel of a linear operator. In other words, we
can find an $M$ by $N$ matrix 
${\mathbb C}= \{ C_{ij}\}_{i=1\dots M,\, j=1\dots N}$, with 
$C_{ij} = 0,1$, and $M=N-L$, such that 
\begin{eqnarray}
{\cal C} = \{ \ux^{(\alpha)}:\alpha = 1,\dots,2^L\} =
\{\ux\in \{ 0,1\}^N : {\mathbb C}\ux = \u0 \, ({\rm mod}\, 2) \}\, .
\end{eqnarray}
The condition ${\mathbb C}\ux = \u0 \, ({\rm mod}\, 2)$ can be
regarded as a set of $M$ linear equations 
(called {\it constraints} or {\it parity checks})
of the form:
\begin{eqnarray}
C_{i1}x_1+C_{i2}x_2+\dots+C_{iN}x_N = 0\; ({\rm mod}\, 2)\, ,
\label{Constraint}
\end{eqnarray}
with $i = 1, \dots,M$.

To each bit $x_i$, $i = 1,\dots, N$, we assign an {\it a priori}
probability distribution $p_i(x_i)$. 
In the information-theory context, the {\it a priori}
distributions $p_i(x_i)$ are induced by the observation of the
channel output, and by the knowledge of the statistical properties of
the channel.
We are interested in studying the
induced probability distribution over the codewords $\ux^{(\alpha)}$.
In other words we want to consider the following probability
distribution over the strings $\ux$ of $N$ bits:
\begin{eqnarray}
P(\ux) = \frac{1}{Z}\, \delta[{\mathbb C}\ux]\, \prod_{i=1}^N p_i(x_i)\, ,
\label{BitwiseModel}
\end{eqnarray}
where $Z$ is a normalization constant; $\delta[\uz]= 1$ if
$\uz = \u0\, ({\rm mod}\, 2)$, and $\delta[\uz]= 0$ otherwise.

\begin{figure}
\centerline{
\epsfig{figure=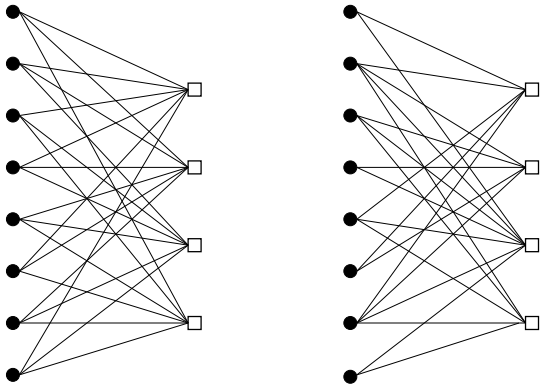,angle=0,width=0.4\linewidth}}
\caption{Two Tanner graphs: a regular one with $(k,l) = (6,3)$ on the
left, and an irregular one on the right. In both cases $N=8$, $M=4$ 
(and therefore the rate is $R=1/2$).}
\label{TannerFigure}
\end{figure}

There are several graphical representations of the above model.
The most used in the coding theory community makes use of the
so-called Tanner graph \cite{Tanner}, cf. Fig. \ref{TannerFigure}. 
This is a bipartite graph which is constructed as follows. 
A node on the left is 
associated to each binary variable $x_j$, and a node on the right to
each constraint, i.e. to each linear equation (\ref{Constraint})
with $i=1,\dots,M$. There are therefore
$N$ left nodes ({\it variable} nodes), and $M$ right nodes ({\it
check} nodes). A given check $i$ is connected to the variables $x_j$ which
appear with nonzero coefficient in the corresponding equation 
(\ref{Constraint}).

The model (\ref{BitwiseModel})  has a spin-wise formulation
\cite{KanterSaad_PRL,KanterSaad_Cascading,Saad_MN_PRL,
KanterSaad_FS,SaadRegular,SaadCactus,SaadTighter} which we
shall employ hereafter. We replace any bit sequence $\ux =
(x_1,\dots,x_N)$, with a spin configuration $\us =
(\sigma_1,\dots,\sigma_N)$, where $\sigma_i= (-1)^{x_i}$. 
The constraints (\ref{Constraint}) on the sums of bits $x_i$,
get translated into constraints on the product of spins $\sigma_i$.
These have the form
\begin{eqnarray}
\sigma^{\omega_i}\equiv\prod_{j\in\omega_i} \sigma_j = +1\, ,
\label{BitConstraint}
\end{eqnarray}
where $\omega_i=\{ j\in \{1,\dots,N\}: C_{ij}=1\}$.
The other ingredient of the model are the {\it a priori} probability
distributions $p_i(x_i)$. They can be encoded into properly chosen
magnetic fields: $p_i(x_i) = e^{\beta h_i \sigma_i}/(2\cosh \beta 
h_i)$, with $2 \beta h_i = \log(p_i(0)/p_i(1))$,
where we introduced the inverse temperature $\beta$ for
later convenience.
With these building blocks, we can write down the spin model
equivalent of Eq. (\ref{BitwiseModel}):
\begin{eqnarray}
P(\us) = \frac{1}{Z(\beta)} \, \prod_{j=1}^M
\delta[\sigma^{\omega_j},+1]\, \exp \left(\beta\sum_{i=1}^N
h_i\sigma_i\right)\, ,
\label{SpinwiseModel}
\end{eqnarray}
where $\delta[a,b]$ is the Kronecker delta function.
This can be regarded as a spin model with infinite strength multi-spin
interactions (which enforce $\sigma^{\omega_j}=+1$) and a 
random magnetic field.

Instead of insisting on the motivations for the probabilistic model
(\ref{SpinwiseModel}) coming from coding theory, we shall remark that,
as it stands, it is remarkably general.
Any spin-model hamiltonian $H(\us) = -\sum_{i_1\dots i_p}
J_{i_i\dots i_p}\sigma_{i_1}\dots\sigma_{i_p}$
can be written in the form (\ref{SpinwiseModel}). 
This can be done by introducing the auxiliary
spin variables $\sigma_{i_1\dots i_p}$. 
The Kronecker delta functions in Eq. (\ref{SpinwiseModel}) 
can be used to enforce  
$\sigma_{i_1\dots i_p} = \sigma_{i_1}\dots\sigma_{i_p}$. The
couplings $J_{i_i\dots i_p}$ become magnetic fields acting on the 
variables $\sigma_{i_1\dots i_p}$.

Untill now we have been pretty generic in the presentation of the model.
In order to be more precise, we have to choose the constraint matrix
${\mathbb C}$, and the magnetic fields $\{ h_i \}_{i=1,\dots, N}$.

Following Gallager \cite{Gallager}, we shall take ${\mathbb C}$ 
to be {\it random} and {\it sparse}.
More precisely ${\mathbb C}$ will be constrained to have $k$
non-zero elements for each row and $l$ non-zero elements for each
column (with $l<k$), and not to have two identical 
rows\footnote{Remark that, with this choice, some of the 
parity check equations (\ref{Constraint}) may be linearly dependent.
However, such an event is {\it rare} for $k>l$ \cite{Gallager}.}. 
This choice corresponds to taking the Tanner graph (cf. Fig. 
\ref{TannerFigure}) as a
random bipartite graph, with variable ({\it left}) nodes of fixed
degree $l$, and check ({\it right}) nodes of degree $k$.
We shall choose among the
matrices of this {\it ensemble} with flat probability distribution.
We shall use the pair $(k,l)$ to denote the spin model (or the error-correcting
code) defined by this {\it ensemble} of matrices.  
An important characteristic of the code is its {\it rate}
$R=1-l/k$, which measures the redundancy of the encoded message 
(infact $R=L/N$).

The magnetic fields $h_i$ will be random i.i.d. variables with
probability distribution $p_h(h_i)$. We consider 
$p_h(h_i)$ to be biased towards positive values of $h_i$ (i.e.
$\int\! dh_i \, p_h(h_i)h_i>0$).
We shall refer often to two simple examples: the two-peak distribution
\begin{eqnarray}
p_h(h_i) = (1-p)\delta(h_i-h_0)+p\delta(h_i+h_0)\, ,
\label{BinaryDistribution}
\end{eqnarray}
with $p<1/2$ and $h_0>0$, and the gaussian distribution
\begin{eqnarray}
p_h(h_i) = \frac{1}{\sqrt{2\pi\tilde{h}^2}}
\exp\left\{-\frac{(h_i-h_0)^2}{2\tilde{h}^2}\right\}\, ,
\label{GaussianDistribution}
\end{eqnarray}
with $h_0>0$. 
It can be shown that, if the model describe communication through 
a noisy ``symmetric'' channel, the condition
\begin{eqnarray}
p_h(-h_i)= e^{-2h_i}p_h(h_i)
\label{ChannelFieldDistribution}
\end{eqnarray}
follows.
This implies $h_0 = (1/2)\log (1-p)/p$ for the example 
(\ref{BinaryDistribution}) (which corresponds to a binary symmetric
channel), and $h_0=\tilde{h}^2$ for the example 
(\ref{GaussianDistribution}) (corresponding to a gaussian channel).
Hereafter we shall denote with $\<\cdot\>_h$ and $\<\cdot\>_{\mathbb
C}$ the averages with respect to the magnetic fields $\{ h_i\}$,
and the {\it ensemble} of matrices ${\mathbb C}$.

More details on the model introduced in this Section, and on analogous
examples can be found in Refs. \cite{Turbo1,Turbo2,KanterSaad_PRL,KanterSaad_Cascading,Saad_MN_PRL,KanterSaad_FS,SaadRegular,SaadCactus,SaadTighter}
%
%
\section{The Nishimori line}
\label{NishimoriSection}

Nishimori \cite{Nishimori1,Nishimori2}
showed that the physics of disordered spin models simplifies 
considerably on a particular line in the phase diagram.
In particular, it has been recently shown \cite{Nishimori3} 
that replica symmetry breaking is absent on this line.
The Nishimori line plays a distinguished role in the correspondence 
between error-correcting codes and disordered spin models.
As shown in Refs. \cite{Rujan,Sourlas5}, 
{\it maximum a posteriori symbol probability} (MAP) decoding for a 
given error-correcting code 
is equivalent to computing expectation values on the Nishimori line
of the corresponding  spin model.

In this Section we extend the results concerning the Nishimori line 
to the model (\ref{SpinwiseModel}).
We shall consider a generic magnetic field distribution
$p_h(h_i)$ satisfying Eq. (\ref{ChannelFieldDistribution}). 
In this case the Nishimori line is simply given by $\beta = 1$.
Although the proofs are very
similar to the ones of Refs. \cite{Nishimori2,Nishimori3}, 
we present them for sake of completeness.
Some consequences of the exact results of this Section will be
outlined in Sec. \ref{ReplicaSection}.

Let us start with some convention. 
Notice that there are two sources of disorder in our model
(\ref{BitwiseModel}): the magnetic field $h_i$ (which is determined by
the channel output), and the check matrix ${\mathbb C}$. 
Different ${\mathbb C}$ correspond to different error-correcting
codes. 
In this Section we keep the 
parity check matrix ${\mathbb C}$ fixed, and average uniquely over the
random magnetic fields $\{ h_i\}$, with distribution $p_h(h_i)$.
Our results will remain valid after averaging with
respect to any {\it ensemble} of check matrices ${\mathbb C}$
(i.e. to any {\it ensemble} of codes).
It is convenient to introduce the notation $\delta_{\mathbb C}[\us]$
to denote the product of Kronecker delta functions in
Eq. (\ref{SpinwiseModel}).
In other words $\delta_{\mathbb C}[\us]=1$, if and only if $\us$
satisfies all the parity checks encoded in ${\mathbb C}$, i.e. if the
corresponding string of bits $\ux$ is a codeword.
We assume that the parity check matrix ${\mathbb C}$ selects $2^L =
2^{NR}$ codewords. This means that there are $2^L$ distinct
configurations $\us$, such that $\delta_{\mathbb C}[\us]=1$.
Finally we shall take the distribution of the random fields 
to satisfy the identity (\ref{ChannelFieldDistribution}).

We start by writing down the definition of the (field averaged) free energy 
density $f_{\mathbb C}(\beta)$ for a given parity check matrix ${\mathbb
C}$:
\begin{eqnarray}
-\beta Nf_{\mathbb C}(\beta) = \int_{-\infty}^{+\infty}\!\prod_{i=1}^N
dh_i\,p_h(h_i)\, \log\left\{\sum_{\us}\delta_{\mathbb C}[\us]\, 
e^{\beta\sum_i h_i\sigma_i}\right\}\, .
\label{FreeDefinition}
\end{eqnarray}
Then we notice, following Ref. \cite{Nishimori2}, 
that the integral over the field $h_i$ can be
decomposed into an integral over its absolute value and a sum over its
sign. Using Eq. (\ref{ChannelFieldDistribution}), we get,
for any function  ${\cal O}(h_i)$ 
\begin{eqnarray}
\int_{-\infty}^{+\infty} \! dh_i \, p_h(h_i) {\cal O}(h_i)=
\int_{0}^{+\infty} \! dh_i \, \rho(h_i)\sum_{\tau_i} e^{h_i\tau_i}
{\cal O}(h_i\tau_i)\, ,
\label{Decomposition}
\end{eqnarray}
where $\rho(h_i)$ is given by
\begin{eqnarray}
\rho(h_i) = \frac{p_h(h_i)+p_h(-h_i)}{2\cosh h_i}\, .
\end{eqnarray}
By using the decomposition (\ref{Decomposition}) into the definition
(\ref{FreeDefinition}), we get 
\begin{eqnarray}
-\beta Nf_{\mathbb C}(\beta) = \int_0^{+\infty}\!\prod_{i=1}^N
dh_i\,\rho(h_i)\, \sum_{\ut}e^{\sum_i h_i\tau_i}
\log\left\{\sum_{\us}\delta_{\mathbb C}[\us]\, 
e^{\beta\sum_i h_i\tau_i\sigma_i}\right\}\, .
\label{Free2}
\end{eqnarray}
To be more compact, we shall use hereafter the shorthand
$\<\cdot \>_{\rho}\equiv \int_0^{+\infty}\!\prod_{i=1}^N
dh_i\,\rho(h_i)\, (\cdot)$ for the average over the absolute values of
the fields $\{h_i\}$.

The next step consists in performing a gauge transformation 
$\tau_i\to \sigma'_i\tau_i$, $\sigma_i\to\sigma'_i\sigma_i$.
Because of the constraint term $\delta_{\mathbb C}[\us]$, the free
energy (\ref{Free2}) is not invariant with respect to such a
transformation for a generic choice of $\{\sigma'_i\}$.
However, if $\delta_{\mathbb C}[\us']=1$, i.e. if $\us'$ is a codeword,
then the gauge transformation leaves  invariant the free energy.
We can sum over all such ``allowed'' transformations,
and divide by their number, namely $2^{NR}$, obtaining
\begin{eqnarray}
-\beta Nf_{\mathbb C}(\beta) = \left\< 
\frac{1}{2^{NR}}\sum_{\ut}\sum_{\us'} 
\delta_{\mathbb C}[\us']
e^{\sum_i h_i\tau_i\sigma_i'}
\log\left\{\sum_{\us}\delta_{\mathbb C}[\us]\, 
e^{\beta\sum_i h_i\tau_i\sigma_i}\right\}\right\>_{\rho}\, ,
\label{Free3}
\end{eqnarray}
where the constraint $\delta_{\mathbb C}[\us']$ force the gauge
transformation $\us'$ to be an allowed one.

In Eq. (\ref{Free3}) we wrote the sums over quenched and dynamical
variables in a symmetric form. This allows to derive several exact 
identities for $\beta=1$, where the symmetry is complete. In
particular, let us consider the internal energy per spin 
$\epsilon_{\mathbb C}(\beta) = \partial_{\beta} (\beta f_{\mathbb
C}(\beta))$. From Eq. (\ref{Free3}) we get
\begin{eqnarray}
\epsilon_{\mathbb C}(\beta = 1)= -\left\< 
\frac{1}{2^{NR}}\sum_{\ut}\sum_{\us}
\delta_{\mathbb C}[\us]\left(\frac{1}{N}\sum_{i=1}^Nh_i\tau_i\sigma_i\right)
e^{\sum_i h_i\tau_i\sigma_i}\right\>_{\rho}\, .
\end{eqnarray}
We can now perform a second gauge transformation $\tau _i\to\tau_i\sigma_i$,
sum over the $\{\sigma_i\}$ using the constraint, and finally sum over
the $\tau_i$. We obtain $\epsilon_{\mathbb C}(\beta = 1) = -\<h\tanh
h\>_h$.
Analogously to Ref. \cite{Nishimori2},  we can further simplify this result,
obtaining
\begin{eqnarray}
\epsilon_{\mathbb C}(\beta = 1) = -\<h\>_h\, ,
\label{NishimoriEnergy}
\end{eqnarray}
which is the first important result of this Section.

We want now to prove the absence of replica symmetry breaking
on the Nishimori line of our model (\ref{BitwiseModel}), i.e. for
$\beta = 1$.
As in Ref. \cite{Nishimori3}, we consider the magnetization distribution 
\begin{eqnarray}
P^{(1)}_{\beta,{\mathbb C}} (m)\equiv \int_{-\infty}^{+\infty}\!\prod_{i=1}^N
dh_i\,p_h(h_i) \frac{\sum_{\us}\,\delta_{\mathbb C}[\us]\,
e^{\beta\sum_i h_i\sigma_i}\,
\delta(m-N^{-1}\sum_i\sigma_i)}{\sum_{\us}\delta_{\mathbb C}[\us]\,
e^{\beta\sum_i h_i\sigma_i}} \, , 
\label{MagnetizationDistribution}
\end{eqnarray}
and the overlap distribution 
\begin{eqnarray}
P^{(2)}_{\beta,{\mathbb C}} (q) \equiv 
\int_{-\infty}^{+\infty}\!\prod_{i=1}^N
dh_i\,p_h(h_i) \frac{\sum_{\us,\us'}\,\delta_{\mathbb C}[\us]\,
\delta_{\mathbb C}[\us']\, 
e^{\beta\sum_i h_i\sigma_i+\beta\sum_i h_i\sigma'_i}\,
\delta(q-N^{-1}\sum_i\sigma_i\sigma_i')}
{\sum_{\us,\us'}\,\delta_{\mathbb C}[\us]\,
\delta_{\mathbb C}[\us']\, 
e^{\beta\sum_i h_i\sigma_i+\beta\sum_i h_i\sigma'_i}}\, .\nonumber\\
\label{OverlapDistribution}
\end{eqnarray}
As before, we keep the parity check matrix ${\mathbb C}$ fixed. 
We shall prove that the two probability distributions defined
above are indeed identical on the Nishimori line $\beta=1$,
i.e. $P^{(1)}_{1,{\mathbb C}} (x)=P^{(2)}_{1,{\mathbb C}} (x)$. Since
the probability distribution of the magnetization 
is expected to be a single delta 
function\footnote{Notice that our model (\ref{BitwiseModel}) 
has no spin-reversal symmetry.} \cite{SpinGlass}, 
this implies the absence of replica symmetry breaking for $\beta = 1$.

We begin by using the decomposition (\ref{Decomposition}) in 
Eq. (\ref{MagnetizationDistribution}). This yields:
\begin{eqnarray}
P^{(1)}_{\beta,{\mathbb C}}(m)= \left\< 
\sum_{\ut}e^{\sum_i h_i\tau_i}
 \frac{\sum_{\us}\,\delta_{\mathbb C}[\us]\,
e^{\beta\sum_i h_i\tau_i\sigma_i}\,
\delta(m-N^{-1}\sum_i\sigma_i)}{\sum_{\us}\delta_{\mathbb C}[\us]\,
e^{\beta\sum_i h_i\tau_i\sigma_i}} \right\>_{\rho}\, . 
\end{eqnarray}
Then we notice that the above distribution is invariant under an
``allowed'' gauge transformation $\tau_i\to\sigma'_i\tau_i$,
$\sigma_i \to\sigma'_i\sigma_i$. As before, ``allowed'' means that 
$\delta_{\mathbb C}[\us']=1$. We can therefore average over these 
transformations, obtaining
\begin{eqnarray}
P^{(1)}_{\beta,{\mathbb C}}(m)= 
\left\< \sum_{\ut,\us'}\delta_{\mathbb C}[\us']
e^{\sum_i h_i\tau_i\sigma_i'}
 \frac{\sum_{\us}\,\delta_{\mathbb C}[\us]\,
e^{\beta\sum_i h_i\tau_i\sigma_i}\,
\delta(m-N^{-1}\sum_i\sigma_i\sigma_i')}{2^{NR}
\sum_{\us}\delta_{\mathbb C}[\us]\,
e^{\beta\sum_i h_i\tau_i\sigma_i}}\right\>_{\rho}\, .\nonumber\\
\end{eqnarray}
We then insert 
$1=(\sum_{\wus}\delta_{\mathbb C}[\wus]e^{\sum_ih_i\tau_i\ws_i})/
(\sum_{\us'}\delta_{\mathbb C}[\us']e^{\sum_ih_i\tau_i\sigma'_i})$,
perform a second gauge transformation $\tau_i\to\ws_i\tau_i$,
$\sigma_i\to\ws_i\sigma_i$, $\sigma'_i\to\ws_i\sigma'_i$, and sum over
$\wus$. Finally we set $\beta = 1$, obtaining 
$P^{(1)}_{1,{\mathbb C}}(m)=P^{(2)}_{1,{\mathbb C}}(m)$, as
anticipated above.
%
%
\section{The random codeword limit}
\label{RandomCodewordSection}

The limiting case $k,l\to\infty$, with $l/k = 1-R$ fixed, plays an
important role. 
We shall call it the random codeword limit for reasons which will be
clear later.
It is a non-trivial limit since the redundancy of
the error-correcting code is kept fixed.
From a theoretical point of view, it allows a simple solution of the
model without changing its qualitative features. 
Our methods will be similar to the ones used by Derrida
to solve the REM \cite{DerridaREM}.
Finally, we will show that
the corrections for finite values of $k$ and $l$ are exponentially
small in $k$. Therefore this limit is interesting also from a
quantitative point of view.

\subsection{The limit $k,l\to\infty$}

Let us consider the probability for a given sequence of bits 
$\ux = (x_1,\dots,x_N)$ to be a codeword with respect to the
{\it ensemble} of parity check matrices ${\mathbb C}$. 
This coincides with the probability $P_{\us}$ for a given spin
configuration $\us$ to satisfy the constraints (\ref{BitConstraint}).
In other words:
\begin{eqnarray}
P_{\us} \equiv \frac{1}{{\cal N}_{\mathbb C}}\sum_{\mathbb C} 
\prod_{j=1}^M \delta[\sigma^{\omega_j},+1]\, ,
\label{OneCodewordProbability}
\end{eqnarray}
where the sum over ${\mathbb C}$ runs over all the matrices 
of the $(k,l)$-{\it ensemble} , and ${\cal N}_{\mathbb C}$ is their number.

Clearly $P_{\us}$ depend upon $\us$ uniquely through the magnetization 
$m_{\sigma}\equiv (1/N)\sum_i \sigma_i$. In general it has the form
\begin{eqnarray}
P_{\us}\sim \exp\left[ N\Sigma_1^{(k,l)}(m_{\sigma}) \right]\, .
\label{OCWresult1}
\end{eqnarray}
The function $\Sigma_1^{(k,l)}(m)$ is computed in Appendix 
\ref{CWProbAppendix} for general
values of $k$ and $l$, and is not particularly illuminating. 
However, in the limit $k,l\to \infty$, $l/k = 1-R$ fixed, we have
\begin{eqnarray}
\Sigma_{(k,l)}(m)\to -(1-R)\log 2\, ,
\label{OCWresult2}
\end{eqnarray}
for any $-1<m<1$. In other words any spin configuration $\us$ has the
same probability $P_{\us} \sim 2^{-(1-R)N}$ of being a codeword.
In addition we must keep track of the 
completely ordered configurations $\sigma_i=+1$ for $i=1,\dots, N$,
and  $\sigma_i=-1$ for $i=1,\dots, N$. The positive one satisfies the
all constraints for any $k$ and $l$, and for any matrix ${\mathbb C}$ 
(this configuration is quite important for the thermodynamics of the model).
The negative one satisfies the constraints for $k$ even, but it is
irrelevant for the thermodynamics.

Let us now turn to a slightly more complicated quantity.
We consider the joint probability $P_{\us,\ut}$ for two different spin 
configurations
$\ut$ and $\us$ to satisfy the same set of constraints (\ref{BitConstraint}),
corresponding to some matrix ${\mathbb C}$ taken from the 
$(k,l)$-{\it ensemble}.
In formulae:
\begin{eqnarray}
P_{\us,\ut} = \frac{1}{{\cal N}_{\mathbb C}}\sum_{\mathbb C}
\prod_{j=1}^M \delta[\sigma^{\omega_j},+1]\delta[\tau^{\omega_j},+1]\,
.\label{TwoCodewordProbability}
\end{eqnarray}
As before we can argue that $P_{\us,\ut}$ depends upon $\us$ and $\ut$
only through their magnetizations $m_{\sigma}$, $m_{\tau}$, and their
overlap $q \equiv (1/N)\sum_i \sigma_i\tau_i$.
The form of $P_{\us,\ut}$ in the thermodynamic limit is
\begin{eqnarray}
P_{\us,\ut} \sim \exp[N
\Sigma_2^{(k,l)}(m_{\sigma},m_{\tau},q)]\, .
\label{TCWresult1}
\end{eqnarray}
The function $\Sigma_2^{(k,l)}(m_1,m_2,q)$ is computed in
Appendix \ref{CWProbAppendix}. 
Again, we shall not report here the result, but we remark 
that in the $k,l\to\infty$ limit
\begin{eqnarray}
\Sigma_2^{(k,l)}(m_1,m_2,q) \to -2(1-R)\log 2\, ,
\label{TCWresult2}
\end{eqnarray}
for any $-1<m_1,m_2,q<1$. In other words, the probability 
for two configurations $\us$, and $\ut$ to satisfy the same set of
constraints is $P_{\us,\ut}\sim P_{\us}P_{\ut}\sim 2^{-2(1-R)N}$:
the two configurations can be regarded as independent ones.

\subsection{The random codeword model}

The previous considerations allow us to replace (in the $k,l\to\infty$
limit) the original model (\ref{SpinwiseModel}), with the following
{\it random codeword model} (RCM).
The model has $2^{NR}$ possible states which we shall index with the
letter $\alpha= 1,\dots, 2^{NR}$. To each of these states we associate
a {\it random} spin configuration 
$\us^{(\alpha)}=(\sigma^{(\alpha)}_1,\dots,\sigma^{(\alpha)}_N)$.
By {\it random} we mean that each spin $\sigma_i^{(\alpha)}$ is
chosen independently from the others, and that 
$\sigma_i^{(\alpha)}=+1$ or $-1$ with equal probability.
Let us underline that, in the random codeword model, the $\sigma^{(\alpha)}_i$
are quenched variables, the dynamical one being the index $\alpha$.
There is a second set of quenched variables: the magnetic fields
$h_i$, with $i = 1, \dots, N$. As in the original model we take them
to be random i.i.d. variables with distribution $p_h(h_i)$.
The energy of the state $\alpha$ reads
\begin{eqnarray}
E^{(\alpha)} = -\sum_{i=1}^N h_i\sigma_i^{(\alpha)}\, .
\label{ERandomWord}
\end{eqnarray}
To the $2^{NR}$ ``disordered'' states described above we add the ordered
state $\alpha = 0$, and the corresponding spin configuration
$\us^{(0)}$, with $\sigma^{(0)}_i  = +1$ for $i=1, \dots, N$.
This corresponds to the ``all zeros'' codeword $\u0$.
Its energy is obviously $E^{(0)} = -\sum_i h_i$.

The random codeword model can be solved through elementary methods. 
Here we shall solve it
for the $\pm h_0$ distribution of fields, see Eq. (\ref{BinaryDistribution}). 
At the end of this Section, we shall quote the result for a general 
distribution $p_h(h_i)$. For sake of clarity we shall report the 
calculation for this case, which is slightly less straightforward,
in the Appendix \ref{RCMAppendix}.

We begin by taking into account the ``random'' states $\alpha =
1,\dots, 2^{NR}$. Later we shall consider the contribution coming from
the ordered state $\alpha = 0$.
Let us consider a fixed configuration of the magnetic fields $\{
h_i\}$. Since the probability distribution of the $\sigma^{(\alpha)}_i$
is flat, $P(\{ \sigma^{(\alpha)}_i \}) = 2^{-N^2R}$, we can apply a
gauge transformation $\sigma^{(\alpha)}_i\to \varepsilon_i
\sigma^{(\alpha)}_i$, with $\varepsilon_i=\pm 1$, without
changing their statistical properties. 
If we choose $\varepsilon_i={\rm sign}(h_i)$, the energy
(\ref{ERandomWord}) becomes $E^{(\alpha)} = -h_0\sum_i
\sigma^{(\alpha)}_i$. We conclude that, for what concerns the ``random''
states, the $\pm h_0$ field distribution is equivalent to an uniform
field $h_i = h_0$.

\begin{figure}
\centerline{\hspace{-1.5cm}
\epsfig{figure=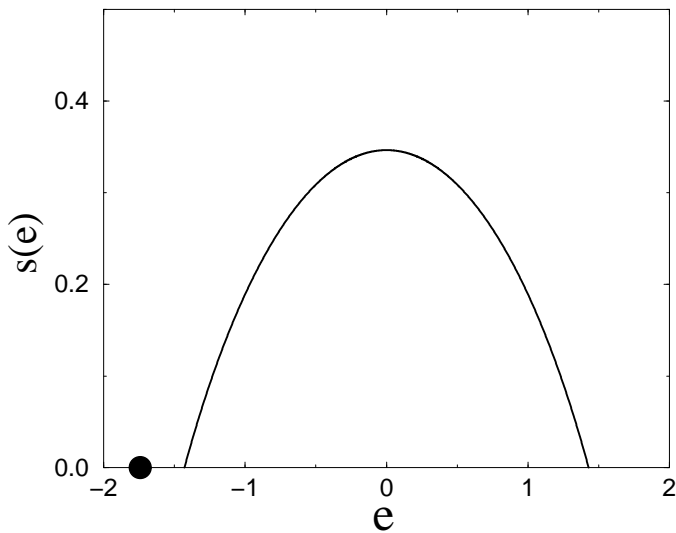,angle=-90,width=0.5\linewidth}}
\caption{The microcanonical entropy density of the RCM
with binary field distribution, cf. Eq. (\ref{BinaryDistribution}).
Here we set $R=1/2$, $p = 0.025$, $h_0 = \atanh(1-2p)$. Notice the 
continuous contribution coming from the random configurations (solid line),
and the isolated ordered configuration (filled circle).}
\label{RcmSpectrum}
\end{figure}
Now we would like to compute the {\it typical} number 
${\cal N} _{typ}(\epsilon)$ of states having
a given energy density $E^{(\alpha)}/N = \epsilon$. This is equal to
the typical number of states having magnetization 
$m^{(\alpha)} = -\epsilon/h_0$. This is a very simple problem.
Define the function 
\begin{eqnarray}
{\cal H}(x) = -\frac{1+x}{2}\log(1+x)-\frac{1-x}{2}\log(1-x)\, .
\label{EntropyDefinition}
\end{eqnarray}
Then
${\cal N} _{typ}(\epsilon) \sim \exp\{ N R\log2 + N {\cal H}(\epsilon/h_0) \}$,
when $|\epsilon|< \epsilon_c$,  and  ${\cal N} _{typ}(\epsilon) = 0$
otherwise. The critical energy $\epsilon_c = h_0
\widehat{\epsilon}(R)$ is the positive solution
of $R\log2 + {\cal H}(\epsilon/h_0) = 0$. The entropy density of the system
$s(\epsilon) = \log {\cal N} _{typ}(\epsilon)/N$ is depicted in Fig. 
\ref{RcmSpectrum}.
Since $s'(-\epsilon_c)>0$ the (sub)system of the random codewords 
undergoes a freezing phase
transition at the critical temperature $\beta_c = s'(-\epsilon_c)$.
This phase transition is analogous to the one of the REM \cite{DerridaREM}: 
it separates an high--temperature paramagnetic phase from
a low--temperature frozen one.

Let us now consider the ordered state $\alpha = 0$, whose energy is 
given by $E^{(0)} = -\sum_i h_i$. In this case we can apply the
central limit theorem. For $N\to\infty$ the 
energy density of the state $\alpha = 0$
is $\epsilon^{(0)} = -(1-2p) h_0$ with probability one. 
We have therefore the following picture of the energy
spectrum of the model: a single ordered state at  $\epsilon^{(0)} = -(1-2p)
h_0$, plus a bell-shaped continuum between $-\epsilon_c(h_0)$ and
$\epsilon_c(h_0)$. The ordered state is thermodynamically
relevant as long as it is separated by a gap from the continuum.
This happens if $p < p_c(R)$, where $p_c(R)$ is the unique solution
between $0$ and $1/2$ of the equation
\begin{eqnarray}
R \log2 + {\cal H}(1-2p)=0\, . 
\label{BinaryCapacity}
\end{eqnarray}
Notice that Eq. (\ref{BinaryCapacity}) coincide with the 
equation determining the capacity of the binary symmetric channel
\cite{Cover}. This means that, in the $k,l\to\infty$ limit, Gallager codes
saturate Shannon capacity.

The free energy is easily determined from the entropy:
\begin{eqnarray}
f(\beta) = \min_{\epsilon}\left\{\epsilon - \frac{1}{\beta}s(\epsilon)
\right\}\, .
\end{eqnarray}
The phase diagram includes three different phases:
a paramagnetic (P) and a spin-glass (SG) phases, associated with the
continuum part of the energy spectrum; a ferromagnetic (F) phase,
associated with the ordered state.
The free energy of the paramagnetic phase is given by:
\begin{eqnarray}
f_P(\beta) = -\frac{R}{\beta}\log 2-\frac{1}{\beta}\log \cosh \beta h_0\, .
\end{eqnarray}
The paramagnetic-spin glass phase boundary is given by the
zero-entropy condition $\partial f_P/\partial \beta = 0$.
We obtain the curve $\beta h_0 = \atanh(1-2p_c(R))\equiv h^*(R)$.
At the transition the system freezes and the free energy in the
spin-glass phase is
\begin{eqnarray}
f_{SG}(\beta) = f_P(\beta = h^*(R)/h_0) = -h_0(1-2p_c(R))\, .
\end{eqnarray}
The ferromagnetic free energy is nothing but the energy of the
ferromagnetic state:
\begin{eqnarray}
f_F(\beta) = -h_0 (1-2p)\, .
\end{eqnarray}
The ferromagnetic-spin glass phase boundary has therefore the simple
form $p=p_c(R)$.

\begin{figure}
\begin{tabular}{cc}
\hspace{-0.5cm}\epsfig{figure=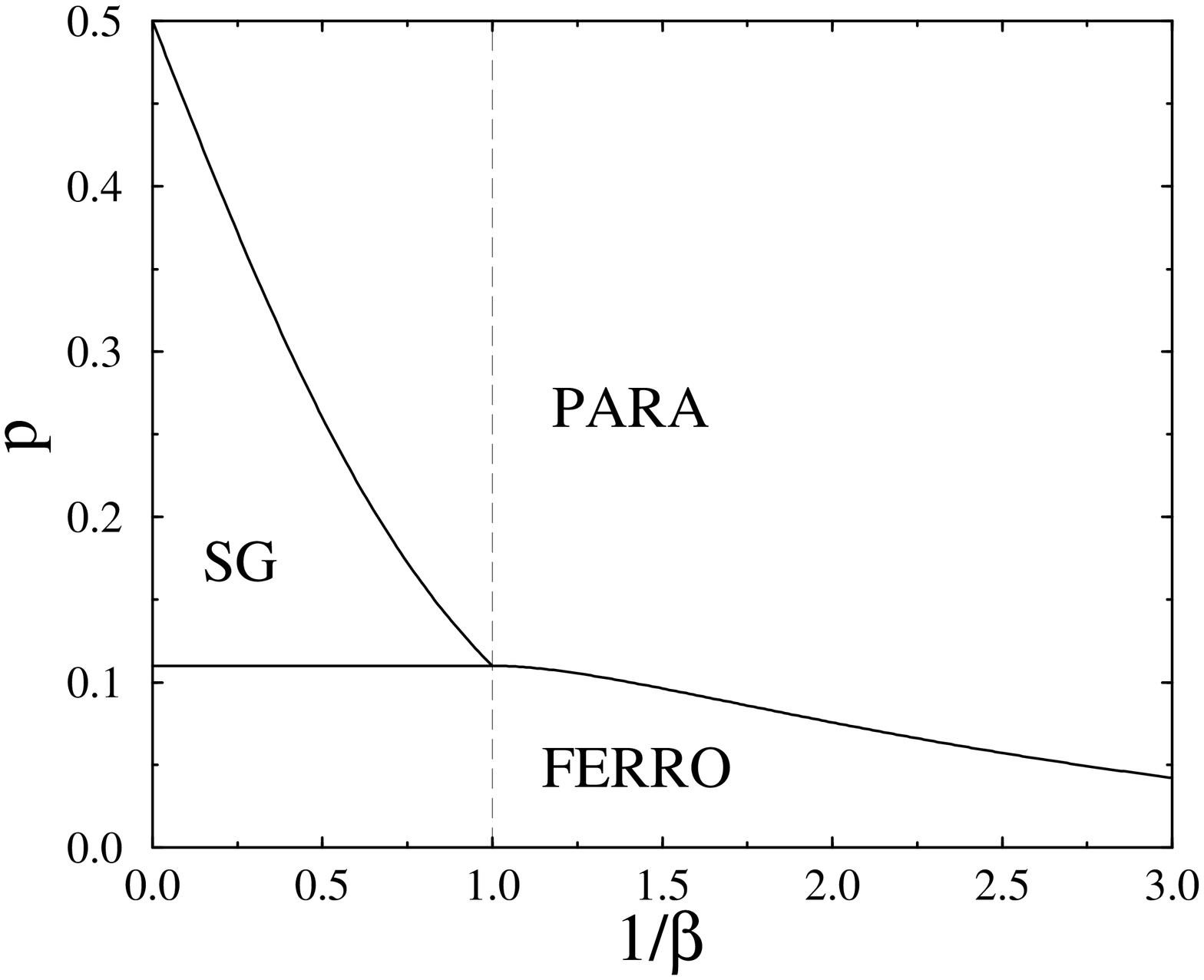,angle=-90,
width=0.5\linewidth}&
\epsfig{figure=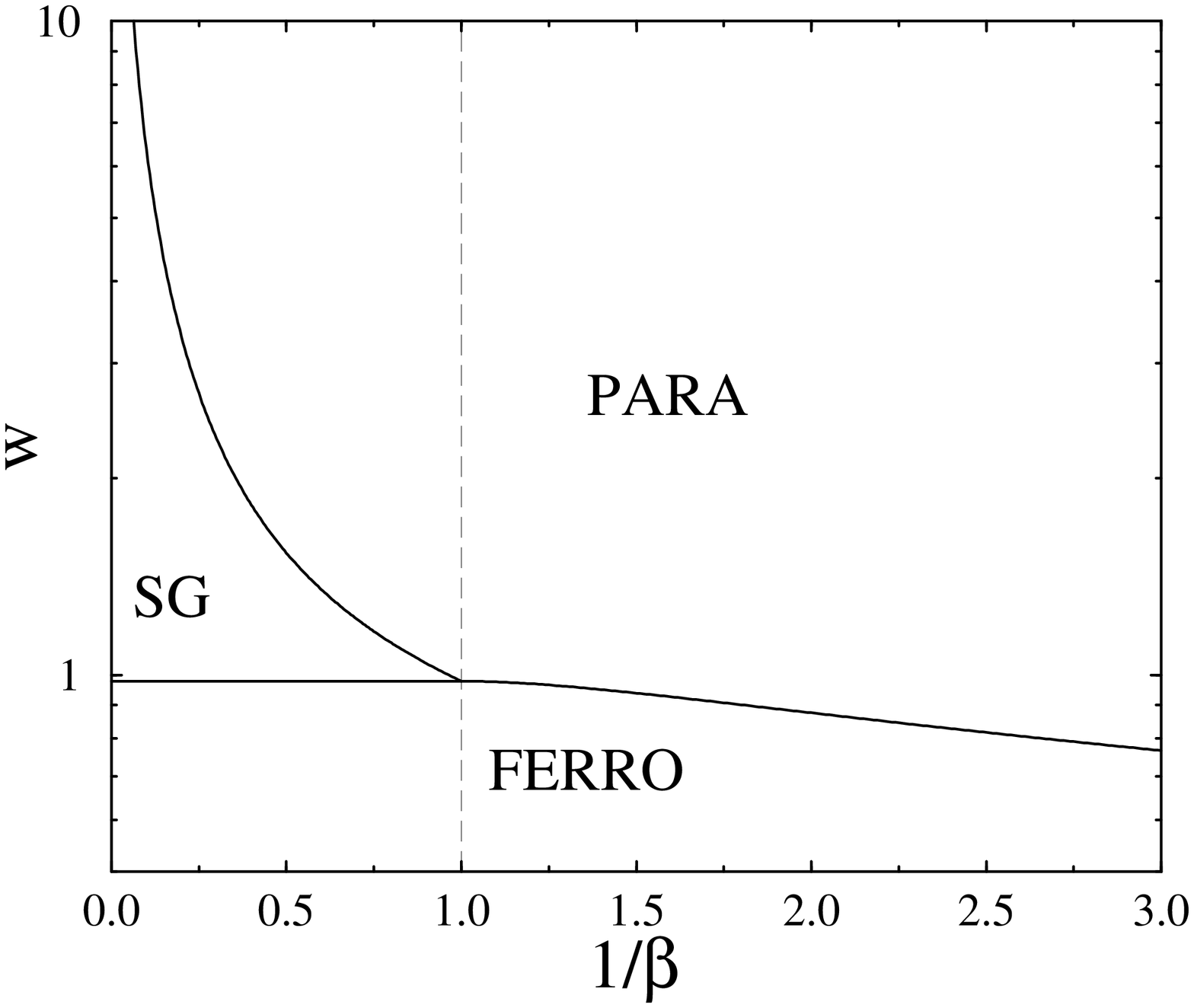,angle=-90,
width=0.48\linewidth}
\end{tabular}
\caption{The phase diagram for binary (left, see Eq. 
(\ref{BinaryDistribution})), and gaussian (right, see Eq. 
(\ref{GaussianDistribution})) field distribution.
In both cases the field distribution was chosen to satisfy 
Eq. (\ref{ChannelFieldDistribution}).}
\label{BscGaussianPhases}
\end{figure}
For sake of clarity, let us consider 
the magnetic field distribution which describes a binary symmetric
channel, i.e. let us fix $h_0 = h_0(p) \equiv \atanh(1-2p)$, cf. Eq.
(\ref{ChannelFieldDistribution}).
The resulting phase diagram is reported in Fig. \ref{BscGaussianPhases}. 
The ferromagnetic-spin glass phase
boundary is at $p=p_c(R)$. The paramagnetic-spin glass boundary is
$\beta \, \atanh(1-2p) = \atanh(1-2p_c(R))$. Finally the 
ferromagnetic-paramagnetic phase boundary is given by
\begin{eqnarray}
R\log 2+\log\cosh\beta h_0(p) -\beta h_0(p)\tanh h_0(p) = 0\, .
\end{eqnarray}
The triple point is at $\beta = 1$, $p = p_c(R)$, and lies
on the Nishimori line.

Untill now we treated the simple case of a two-peak distribution of
the magnetic fields: $p_h(h_i) =
(1-p)\, \delta(h_i-h_0)+p\, \delta(h_i+h_0)$. What does it happen for a generic
$p_h(h_i)$?
In Appendix \ref{RCMAppendix} it is shown that the same scenario applies with
some slight modification. The free energy in the paramagnetic phase 
becomes
\begin{eqnarray}
f_P(\beta) = -\frac{R}{\beta}\log 2-\frac{1}{\beta}\<\log \cosh \beta
h\>_h\, .
\label{FreeRCWGeneric}
\end{eqnarray}
The system undergoes a freezing transition at a critical temperature
$\beta_c$ determined from the condition 
$\left.\partial f/\partial \beta\right|_{\beta_c} = 0$.
For $\beta >\beta_c$, the system is in a glassy phase with free energy 
$f_{SG}(\beta) = f_{P}(\beta_c)$.
Finally, the ferromagnetic phase coincides with the ordered state
$\alpha = 0$, and has free  energy $f_F(\beta) = -\< h\>_h$. 

To be specific we report in Fig. \ref{BscGaussianPhases} 
the phase diagram for the gaussian distribution
\begin{eqnarray}
p_h(h) = \sqrt{\frac{w^2}{2\pi}}
\exp\left\{-\frac{w^2}{2}\left[h-\frac{1}{w^2}\right]^2\right\}\, ,
\end{eqnarray}
which describes a gaussian channel with noise variance $w$.
The triple point is located at $\beta = 1$ and $w = w_c(R)$,
$w_c(R)$ being the solution of the equation below
\begin{eqnarray}
R\log 2+\< \log\cosh h\>_h - \< h\tanh h\>_h = 0\, .
\end{eqnarray}
It is easy to show that the solution $R(w)$ of the above equation
correspond to the capacity of a gaussian channel with constrained
binary inputs \cite{Viterbi}.
%
%
\section{The replica calculation} 
\label{ReplicaSection}

As always \cite{SpinGlass}
we compute the integer moments $\<Z^n\>_{h,{\mathbb C}}$ of
the partition function by replicating the system $n$ times.
To the leading exponential order we get 
\begin{eqnarray}
\<Z^n\>_{h,{\mathbb C}} \sim \int \!\prod_{\ps} d\l(\ps)d\lh(\ps)
\, e^{-N S[\l,\lh]}\, ,
\label{Zn}
\end{eqnarray}
where
\begin{eqnarray}
S[\l,\lh] & = & l\sum_{\ps} \l(\ps)\lh(\ps)-
\frac{l}{k}\sum_{\ps_1,\dots,\ps_k} \l(\ps_1)\cdot\dots\cdot\l(\ps_k)
\prod_{a=1}^n \delta[\sigma^a_1\dots\sigma^a_k,+1]-\nonumber \\
&&-\log\left\{\sum_{\ps}\lh(\ps)^l\<e^{\beta h\sum_a\sigma^a}\>_h\right\}
-l+\frac{l}{k}\, ,
\label{ReplicatedAction}
\end{eqnarray}
and $\ps = (\sigma^1,\dots,\sigma^n)$ is the replicated spin variable.
The calculations which lead to Eq. (\ref{ReplicatedAction}) 
are completely analogous to the ones
of Refs. \cite{SaadRegular,SaadTighter}.
To be self-contained we shall sketch them in Appendix \ref{ReplicaAppendix}.
The free energy $f(\beta)$ is obtained by taking the saddle point
of the integral (\ref{Zn}) (let say $\l = \l_n^*$, $\lh =\lh_n^*$) 
and evaluating the $n\to 0$ limit:
$\beta f(\beta) = \lim_{n\to 0} \partial_n S[\l_n^*,\lh_n^*]$.
The saddle point equations are
\begin{eqnarray}
\lh(\ps) & = &
\sum_{\ps_1,\dots,\ps_{k-1}} \l(\ps_1)\cdot\dots\cdot\l(\ps_{k-1})
\prod_{a=1}^n \delta[\sigma^a\sigma^a_1\dots\sigma^a_{k-1},+1]\, ,
\label{Saddle1}\\
\l(\ps) & = & \frac{\lh(\ps)^{l-1}\<e^{\beta h\sum_a\sigma^a}\>_h}
{\sum_{\ps}\lh(\ps)^l\<e^{\beta h\sum_a\sigma^a}\>_h}\, .
\label{Saddle2}
 \end{eqnarray}

The above equations are satisfied by the totally ordered
solution
$\l_0(\ps) =\lh_0(\ps) = \delta_{\ps,\ps_0}$, where $\ps_0=(+1,\dots,+1)$.
The corresponding free energy is $f_F(\beta) = -\<h\>_h$. Such a  solution
is is possible because of the infinite-strength ferromagnetic
interactions in our model (\ref{BitwiseModel}). 
Physically  it is related to the configuration 
$\{\sigma_i=+1\}_{i=1,\dots,N}$, which satisfies all the 
constraints\footnote{Notice that, for $k$ even, there are $2^n$ solutions
of the type $\l(\ps) = \lh(\ps) = \delta_{\ps,\pt}$. The ``spurious''
solutions with $\pt\neq\ps_0$ are related to the  
$\{ \sigma_i=-1\}_{i=1,\dots,N}$ configuration. Since we took 
$\<h\>_h >0$, these solutions do not have thermodynamical relevance.}. 

\subsection{Stability of the ferromagnetic phase}

In the ferromagnetic solution found above (as in the ferromagnetic phase
found in Sec. \ref{RandomCodewordSection}) the system is completely
ordered (i.e. the magnetization is $m=1$). 
This correspond to no-error communication in the coding language.
Knowing the boundaries of the ferromagnetic phase
is therefore of great practical relevance.
Here we shall investigate the issue of local stability.
The calculation is similar (although much simpler) to the one carried
out for turbo codes in Ref. \cite{Turbo2}.

We start by computing the replicated action (\ref{ReplicatedAction})
for $\l(\ps)$, $\lh(\ps)$ ``near'' the ferromagnetic saddle point,
namely $\l(\ps) =\l_0(\ps)+\d(\ps)$, $\lh(\ps) =\lh_0(\ps)+
\dh(\ps)$. We first consider the case $l>2$: 
\begin{eqnarray}
\delta S[\l_0,\lh_0] = l\sum_{\us}\d(\us)\dh(\us)
-\frac{1}{2}l(k-1)\sum_{\us}\d(\us)^2+\frac{1}{2}l\,\dh(\us_0)^2+O(\delta^3)
\, ,
\end{eqnarray}
where $\delta S[\l_0,\lh_0] \equiv S[\l_0+\d,\lh_0+\dh]-S[\l_0,\lh_0]$.
It is convenient to integrate over $\l(\us)$ using the saddle point 
equation (\ref{Saddle1}), which, for $\l(\ps) =\l_0(\ps)+\d(\ps)$, 
$\lh(\ps) =\lh_0(\ps)+\dh(\ps)$, gives 
$\delta(\ps) = \dh(\ps)/(k-1)+O(\delta^2)$. We finally get
\begin{eqnarray}
\delta S[\lh_0] = \frac{1}{2}\sum_{\ps}
\zeta_{\ps}\dh(\ps)^2+O(\d^2)\, ,
\label{QuadraticForm}
\end{eqnarray}
where $\zeta_{\ps_0}=lk/(k-1)$, and $\zeta_{\ps}=l/(k-1)$ for
$\ps\ne\ps_0$. We conclude that, for $l>2$, the ferromagnetic phase is
always locally stable and its boundaries must correspond to first
order phase transitions.

For $l=2$ the situation is physically different. Equation 
(\ref{QuadraticForm}) is still valid,
with $\zeta_{\ps_0}=2k/(k-1)$ and
\begin{eqnarray}
\zeta_{\ps} = 2\left[ \frac{1}{k-1}-\frac{\<e^{\beta
h\sum_a\sigma^a}\>_h}{\<e^{\beta h n}\>_h}\right]
\end{eqnarray}
for $\ps\ne \ps_0$. We have therefore $n$ different eigenvalues
$\zeta_{n,\omega}$, with degeneracies $\left(
\begin{array}{c} n\\ \omega\end{array}\right)$, where
$\omega\equiv n-\sum_a\sigma^a$. The first instability occurs for $\omega=1$.
The corresponding critical line is given by $(k-1)\<e^{-\beta_ch}\>_h
=1$.
This local stability condition is already known 
\cite{RichardsonUrbanke} in the coding
community, although it has been obtained by completely different methods.

Hereafter we shall focus on the case $l\ge 3$.

\subsection{Replica symmetric approximation}

The simplest approximation for treating the $n\to 0$ limit, consists
in choosing $\l(\ps)$ and $\lh(\ps)$ to be replica symmetric, i.e. to
depend upon $\ps$ uniquely through the symmetric combination 
$\sum_a\sigma^a$. A commonly adopted parametrization \cite{Wong} 
is the following
\begin{eqnarray}
\l(\ps) = \int\! dx \, \pi(x) \frac{e^{\beta x\sum_a\sigma^a}}
{(2\cosh \beta x)^n}\, ,
\label{ReplicaSymmetricAnsatz}
\end{eqnarray}
and the analogous one for $\lh(\ps)$ (with a different distribution
$\ph(y)$). The replica symmetric order parameters $\pi(x)$ and 
$\ph(y)$ have the physical meaning of probability distributions of
cavity fields. In particular
\begin{eqnarray}
P(H) = \int\!dx\, \pi(x)\int\!dy\, \ph(y)\, \delta(H-x-y)\, ,
\end{eqnarray}
is the probability distribution of the effective fields
$H_i \equiv (1/\beta) \atanh \<\sigma_i\>$.

Using the ansatz (\ref{ReplicaSymmetricAnsatz}), we easily 
obtain the replica symmetric free energy:
\begin{eqnarray}
\beta f_P[\pi,\ph] & = & \frac{l}{k}\log 2 -\<\log\cosh\beta
h\>_h+ l \int\! dx\, \pi(x)\int\!dy\,\ph(y)\,
\log[1+\t(x)\t(y)]-\nonumber\\
&&-\frac{l}{k}\int\! dx_1\, \pi(x_1)\dots\int\!dx_k\,\pi(x_k)
\log[1+\t(x_1)\dots\t(x_k)]-\nonumber\\
&&-\int\! dy_1\, \ph(y_1)\dots\int\!dy_l\,\ph(y_l)
\< \log {\mathbb F}_l(h,y_1,\dots,y_l;\beta) \>_h\, ,
\label{RSFreeEnergy}
\end{eqnarray}
where we defined $\t(x)\equiv \tanh \beta x$ and
\begin{eqnarray}
{\mathbb F}_l(y_0,y_1,\dots,y_l;\beta)\equiv 
\prod_{i=0}^l(1+\t(y_i))+\prod_{i=0}^l(1-\t(y_i))\, .
\end{eqnarray}
The field distributions $\pi(x)$ and $\ph(y)$ are determined by the
saddle point equations:
\begin{eqnarray}
\ph(y) & = & \int\!dx_1\, \pi(x_1)\dots \int\!dx_{k-1}\,
\pi(x_{k-1})\, 
\delta\left[y-\frac{1}{\beta}\atanh(\t(x_1)\dots\t(x_{k-1}))\right]\,
,\nonumber\\
\label{SaddleRS1}\\
\pi(x) & = &  \int\!dy_1\, \ph(y_1)\dots \int\!dy_{l-1}\, \pi(y_{l-1})
\< \delta(x-h-y_1-\dots-y_{l-1})\>_h\, .
\label{SaddleRS2}
\end{eqnarray}
The above equations can be solved either numerically or in some
particular limit. In the next Section we will see that the expansion
around the random codeword limit provides rather accurate
results.

\subsection{One step replica symmetry breaking}

To go beyond replica symmetric approximation, one has to divide
the $n$ replicas into $n/m$ subgroups of $m$ replicas 
(with $1\le m\le n$). 
The order parameters $\l(\ps)$, and $\lh(\ps)$ depend upon $\ps$
through the $n/m$ variables
$\widehat{\sigma}^{\alpha}\equiv \sum_{a=m(\alpha-1)+1}^{m\alpha}\sigma^a$.
As discussed clearly in Refs. \cite{MonassonRSB,MezardParisiBethe}, 
in the $n\to 0$ limit the  order
parameter becomes a functional over a probability space
and the calculations becomes rather
cumbersome (see Refs. \cite{MezardParisiBethe,BiroliVariational}
for two viable approaches).

In our case there exists a very simple solution to the saddle point
equations (\ref{Saddle1}), (\ref{Saddle2})
incorporating one step replica symmetry breaking:
\begin{eqnarray}
\l(\ps) = \sum_{\{s^{\alpha}\}}
\int\! dx\, \pi_m(x) \frac{e^{\beta x\sum_{\alpha =
1}^{n/m}s^{\alpha}}}
{(2 \cosh\beta x)^{n/m}}
\prod_{\alpha=1}^{n/m}\prod_{a=(\alpha-1)m+1}^{\alpha m} 
\delta[\sigma^a,s^{\alpha}]\, ,
\label{RSBreaking}
\end{eqnarray}
and the analogous one for $\lh(\ps)$ (with a different distribution
$\ph_m(y)$). It is easy to see that the above ansatz satisfies the 
saddle point equations as soon as $\pi_m(x)$, $\ph_m(y)$  are solution
of the replica symmetric equations (\ref{SaddleRS1}),
(\ref{SaddleRS2}), with the substitution $h\to mh$. The phase
described by the solution (\ref{RSBreaking}) is completely analogous
to the spin-glass phase found in the random codeword model. The system 
is frozen in a large number of ``optimal'' 
configurations (with self-overlap $q_{EA} = 1$). The
overlap between two such configurations is 
$q_0 = \int\! dx\,\pi_m(x)\int\! dy\,\ph_m(y)\, \t^2(x+y)$.

Such a simple scenario (and the simple solution (\ref{RSBreaking}))
is possible because the multi-spin interactions
of the model (\ref{SpinwiseModel}) have infinite-strength.
The existence of other replica-symmetry-breaking solutions
is an open issue, see Sec. \ref{DiscussionSection}.
In the next Section we will show that our ansatz
gives back the RCM solution, see Sec. \ref{RandomCodewordSection}, 
in the $k,l\to\infty$ limit.

The free energy of the solution (\ref{RSBreaking}) is
$f_{SG,m}(\beta) = f_P(\beta m)$, see Eq. (\ref{RSFreeEnergy}), 
and has to be optimized over $m$ with
$0\le m\le 1$. This procedure yields the spin-glass free energy
$f_{SG}(\beta) = f_P(\beta_c)$, and $m = \beta_c/\beta$. 
The critical temperature $\beta_c$ is given 
by the marginality condition $\partial_m
f_{SG,m}(\beta)|_{m=1} = 0$, which coincides with the zero-entropy
condition $\partial_{\beta} f_P(\beta)|_{\beta=\beta_c}=0$.

Let us now draw some consequences of our solution (\ref{RSBreaking})
for the phase diagram of the model.
Since both the spin-glass and the ferromagnetic free energies are
temperature independent, the ferromagnetic-spin glass phase boundary
must stay parallel to the temperature axis. If, for instance,
we consider the binary field distribution
(\ref{BinaryDistribution}) with $h_0=\atanh (1-2p)$, this boundary is
simply given by $p = p_c(k,l)$. Moreover we notice that the energy density
on the line $\beta =1$, see Eq. (\ref{NishimoriEnergy}), 
is equal to the ferromagnetic free energy.
This implies that the entropy vanishes at the ferromagnetic-paramagnetic
boundary for $\beta = 1$. Since the paramagnetic-spin glass boundary
is determined by the zero entropy condition, this point must be the
triple point.
In synthesis, the main characteristics of the phase diagram depicted
in Fig. \ref{BscGaussianPhases} remain valid for finite connectivities.
%
%
\section{Large $k,l$ expansion}
\label{ExpansionSection}

Here we show that the replica solution exhibited in the
previous Section goes to the random codeword model solution 
(cf. Sec. \ref{RandomCodewordSection}) when $l,k\to\infty$ at $l/k=1-R$ fixed. 
Moreover we want to stress that this limit
can be useful from a quantitative point of view. In fact, the corrections for
finite $k$ are exponentially small in $k$.

Notice that the free energy in the spin glass phase $f_{SG}(\beta)$ is easily 
obtained from the paramagnetic free energy $f_P(\beta)$.
In fact we have $f_{SG}(\beta) = f_P(\beta_c)$, where the freezing
temperature $\beta_c$ is given by the zero-entropy condition
$\partial_{\beta}f_P(\beta) = 0$.
Moreover the ferromagnetic free energy is $f_F(\beta) = -\<h\>_h$,
and does not depend upon $k$ and $l$.
It is then sufficient to solve Eqs. (\ref{SaddleRS1}),
(\ref{SaddleRS2}) for large $k,l$ and evaluate Eq. 
(\ref{RSFreeEnergy}) on the solution. The result is $f^{(exp)}_P(\beta)$
($exp$ stands for ``expanded''), and allow to reconstruct the whole phase
diagram as explained above.

The expansion is obtained by noticing that the product 
$\t(x_1)\cdot\dots\cdot\t(x_{k-1})$ which appears
on the right-hand side of Eq. (\ref{SaddleRS1}) is exponentially
small in $k$ as long as $\pi(x)$ is supported on finite values of $x$.
We then expand the the right-hand side of Eq. (\ref{SaddleRS2})
for small values of $y$ and plug the result in 
Eq. (\ref{SaddleRS1}).

The calculations are straightforward. 
For sake of simplicity we show some consequences
for the two-peak field distribution (\ref{BinaryDistribution}).
We refer to Appendix \ref{FormulaeAppendix} for the general results.

\begin{figure}
\centerline{
\epsfig{figure=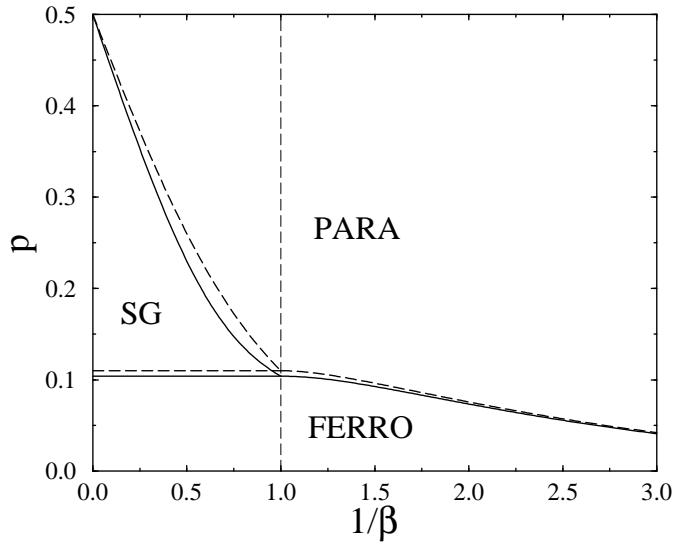,angle=-90,width=0.5\linewidth}}
\caption{The phase diagram for the $(6,3)$ code as computed
from the large $k$, $l$ expansion (continuous lines),
and the one of the RCM (dashed lines). The vertical dashed line is the
Nishimori line $\beta = 1$.}
\label{BscPhasesExp}
\end{figure}
In Fig. \ref{BscPhasesExp} we report the modified phase diagram for 
$k=6$, $l=3$,
as computed using the expansion of Appendix \ref{FormulaeAppendix} 
(cf. Eq. (\ref{FreeExpansion}))
for the paramagnetic free energy. 
We consider the two-peak distribution (\ref{BinaryDistribution})
with $h_0 = \atanh(1-2p)$.
The paramagnetic/spin-glass boundary is obtained by imposing the
zero-entropy condition $\partial_{\beta}f^{(exp)}_P(\beta)= 0$. 
We set $f^{(exp)}_{SG} (\beta) \equiv f^{(exp)}_P(\beta_c)$.
The ferromagnetic spin-glass, and ferromagnetic/paramagnetic
boundaries are obtained by imposing $f_F(\beta) =
f^{(exp)}_{SG}(\beta)$, and $f_F(\beta) = f^{(exp)}_P(\beta)$.

The triple point is at $\beta =1$, $p=p_c(k,l)$.
As we stressed in Sec. \ref{NishimoriSection}, the line $\beta=1$ is of great
practical importance, since it correspond to a widespread decoding 
procedure (MAP decoding).
The critical noise $p_c(k,l)$ has the meaning of the threshold for
no-error communication under MAP decoding.
Since the ferromagnetic-spin glass phase boundary 
stays parallel to the temperature axis, $p_c(k,l)$ is also the threshold
for any ``finite-temperature'' decoding \cite{Rujan} for $\beta\ge 1$.
We get
\begin{eqnarray}
p_c(k,l) = p_c^0-\frac{1-R}{4 {\cal
H}'(1-2 p^0_c)}(1-2p^0_c)^{2k}+O((1-2p^0_c)^{4k})\, ,
\label{AsymptoticThereshold}
\end{eqnarray}
where the function ${\cal H}(x)$ has been defined in Eq. 
\ref{EntropyDefinition}.
In the $k,l\to\infty$ limit, we recover the threshold $p^0_c\equiv p_c(R)$ of
the random codeword model, given by the solution of 
Eq. (\ref{BinaryCapacity}). The deviations from the {\it optimal} properties
of the random-codeword model are exponentially small for large $k$.

\begin{figure}
\centerline{
\epsfig{figure=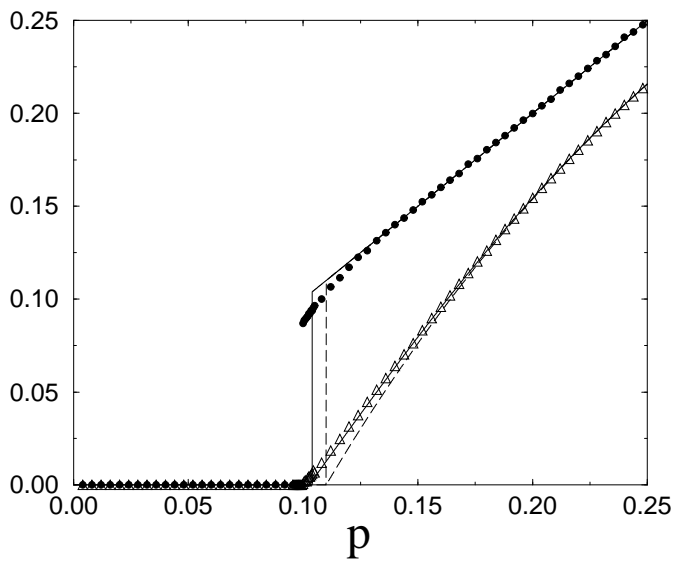,angle=-90,width=0.5\linewidth}}
\caption{The error probability per bit (filled circles and upper curves), 
and the entropy (empty triangles and lower curves) for the $(6,3)$ model with 
binary field distribution (\ref{BinaryDistribution}). 
We set $\beta=1$ and $h_0=\atanh(1-2p)$. The 
symbols are obtained by solving numerically the saddle point 
equations (\ref{SaddleRS1}), (\ref{SaddleRS2}).
The dashed lines are the RCM results. The continuous lines 
are the results of the large-connectivity expansion.}
\label{EprobExp}
\end{figure}
Equations (\ref{SaddleRS1}) and (\ref{SaddleRS2}) can be solved
numerically by a ``population dynamics'' algorithm. One represents the
distributions $\pi(x)$ and $\ph(y)$ by two populations $\{
x_i\}_{i=1,\dots,{\cal L}}$ and $\{y_j\}_{j=1,\dots,{\cal L}}$, and
then iterates the equations (\ref{SaddleRS1}) and (\ref{SaddleRS2}).
This method has been already used, for instance,
in Ref. \cite{MezardParisiBethe}.
In Fig. \ref{EprobExp}, we consider once again the line $\beta = 1$ and
compare the results of large $k,l$ expansion with the numerical
solution of Eqs. (\ref{SaddleRS1}) and (\ref{SaddleRS2}).
We plot both  the entropy and the average error probability per bit
$\<P_e\>_{h,{\mathbb C}}$, where:
\begin{eqnarray}
P_e = \frac{1}{N}\sum_{i=1}^N \frac{1}{2}(1-{\rm sign}\<\sigma_i\>)\,,
\end{eqnarray}

As conclusion let us consider the problem of calculating
the critical noise $p_c(k,l)$. This can be obtained either by 
solving numerically  Eqs. (\ref{SaddleRS1}) and
(\ref{SaddleRS2}), or from the expansion (\ref{AsymptoticThereshold}).
The numerical solution yields $p_c(k,l) = 0.0997(2)$,
$0.1071(2)$, $0.1091(2)$, 
for, respectively, $(k,l) = (6,3)$, $(8,4)$, $(10,5)$. 
From the expansion (\ref{AsymptoticThereshold}) we get $p^{exp}_c(k,l)\approx
0.103965$, $0.107783$, $0.109195$ for the same values of $k$ and $l$.
%
%
\section{Finite size corrections and numerical results}
\label{FiniteSizeSection}

In this Section we compare the analytical predictions with numerical
results in order to confirm the validity of the former and to
investigate the nature of finite size corrections. Needless to say,
the last one is a point of utmost practical importance in coding theory.
Indeed it is known that the thermodynamic limit is approached 
exponentially fast in the ferromagnetic phase, 
at zero temperature \cite{Viterbi}. 
We expect the same behavior to hold in the whole ferromagnetic phase.

Here we focus on the paramagnetic-spin glass phase transition.
We compute the finite size corrections to the free energy of the
RCM. This calculation 
is compared with exact enumeration calculations on small systems.
Then we switch to the complete model (\ref{SpinwiseModel}) and 
compare the the numerical results with the outcome of the replica
calculations, cf. Sec. \ref{ReplicaSection}.

\subsection{The random codeword model}

Let us consider, for sake of clarity, the binary distribution
(\ref{BinaryDistribution}) with $p > p_c(R)$. This corresponds to
focusing on the paramagnetic-spin glass phase transition. Under this
condition the ordered state $\alpha = 0$ belongs to the continuous part 
of the spectrum and there is no energy gap.
We shall therefore neglect this state. Its contribution is
exponentially small in the thermodynamic limit.

With this assumption, we obtain the following result for the free
energy density
\begin{eqnarray}
f(\beta,N) = f_0(\beta) +\frac{1}{N}f_1(\beta,N)+O(1/N^2)\, ,
\label{FreeFS}
\end{eqnarray}
The leading term has been already computed in
Sec. \ref{RandomCodewordSection}. The first correction $f_1(\beta,N)$
vanishes in the paramagnetic phase and depends weakly upon $N$.
Explicit formulae are given in Appendix \ref{FiniteSizeAppendix}.
In particular $f_1(\beta,N)\sim (1/2\beta_c)\log N$ as $N\to\infty$.
The leading correction in the paramagnetic phase is exponentially
small in $N$. In order to compute it, the ferromagnetic state cannot
be neglected.

It is very easy to compute numerically the finite-$N$ free energy
for the random codeword model with binary field distribution
(\ref{BinaryDistribution}), as long as we neglect the ordered state.
All we need, for a given sample, is the energy spectrum. Let us call 
$\nu_k$, with $k = 0,\dots,N$ the number of states $\alpha$,
such that $E^{(\alpha)}= -h_0(N-2k)$. The probability distribution of
the spectrum $\{\nu_k\}$ is
\begin{eqnarray}
P(\{\nu_k\}) = \frac{{\cal N}!}{\prod_{k=0}^N\nu_k !}
\prod_{k=0}^Np_k^{\nu_k}\, ,
\label{NuDistribution}
\end{eqnarray}
where $\sum_k \nu_k = {\cal N} \equiv 2^{NR}$, and
\begin{eqnarray}
p_k \equiv \frac{1}{2^N}\left(\begin{array}{c} N\\ k \end{array}\right)\, .
\end{eqnarray}
Once the $\{\nu_k\}$ have been generated with probability distribution
(\ref{NuDistribution}), the partition function is given by
$Z(\beta) = \sum_k\nu_k\exp\{\beta h_0 (N-2k)\}$.

We considered the RCM with rate $R = 1/2$ and binary field
distribution (\ref{BinaryDistribution}) with 
$h_0 = \atanh(1-2p)$. The phase diagram of this model is depicted in
Fig. \ref{BscGaussianPhases}.
We fixed the flip probability $p = 0.2$ to be greater than the
threshold $p_c(1/2) \approx 0.110025$, and computed the temperature
dependence of the free energy by averaging over $10^5$ realizations of
the spectrum $\{\nu_k\}$.

\begin{figure}
\begin{tabular}{cc}
\hspace{-0.6cm}\epsfig{figure=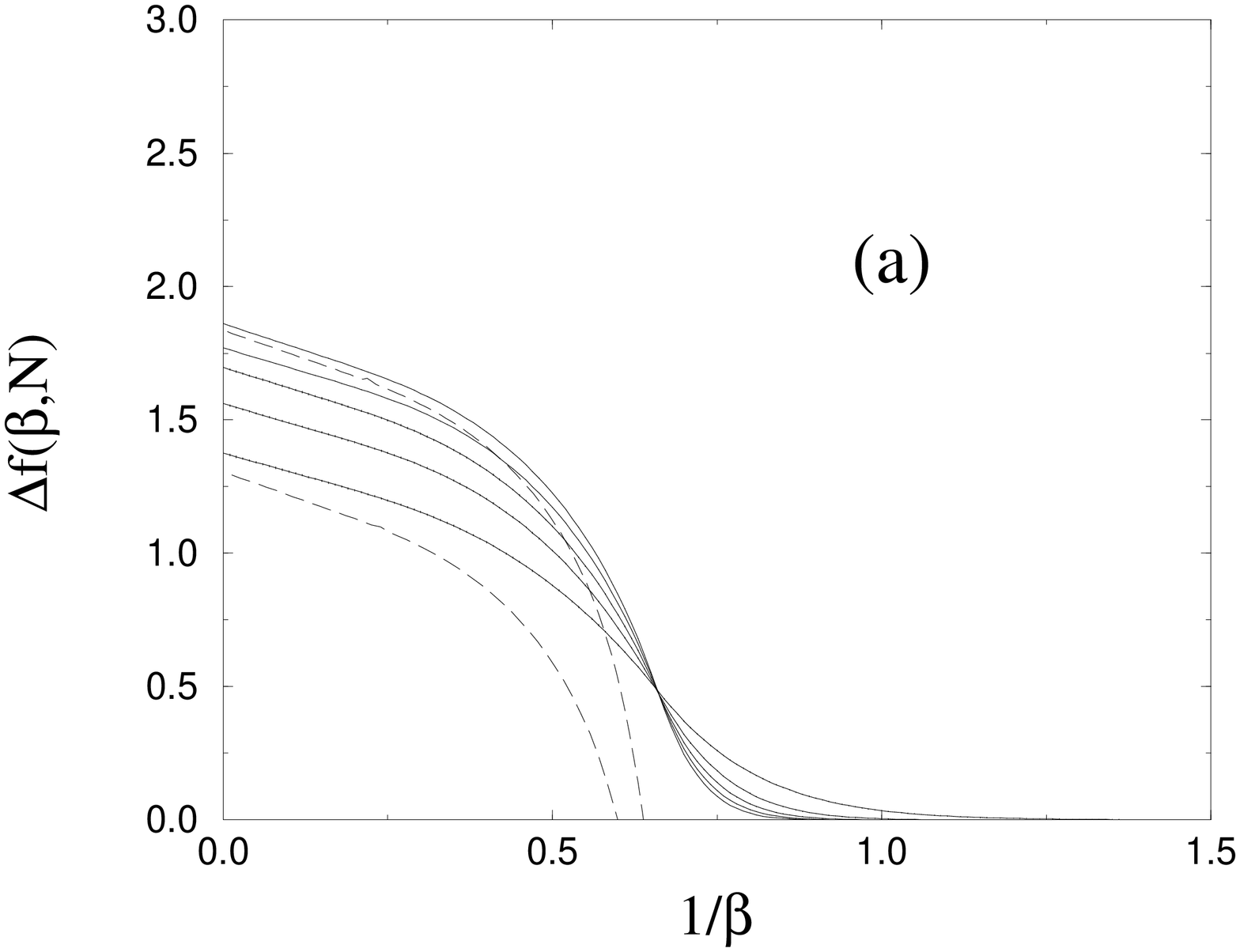,angle=-90,width=0.5\linewidth}&
\epsfig{figure=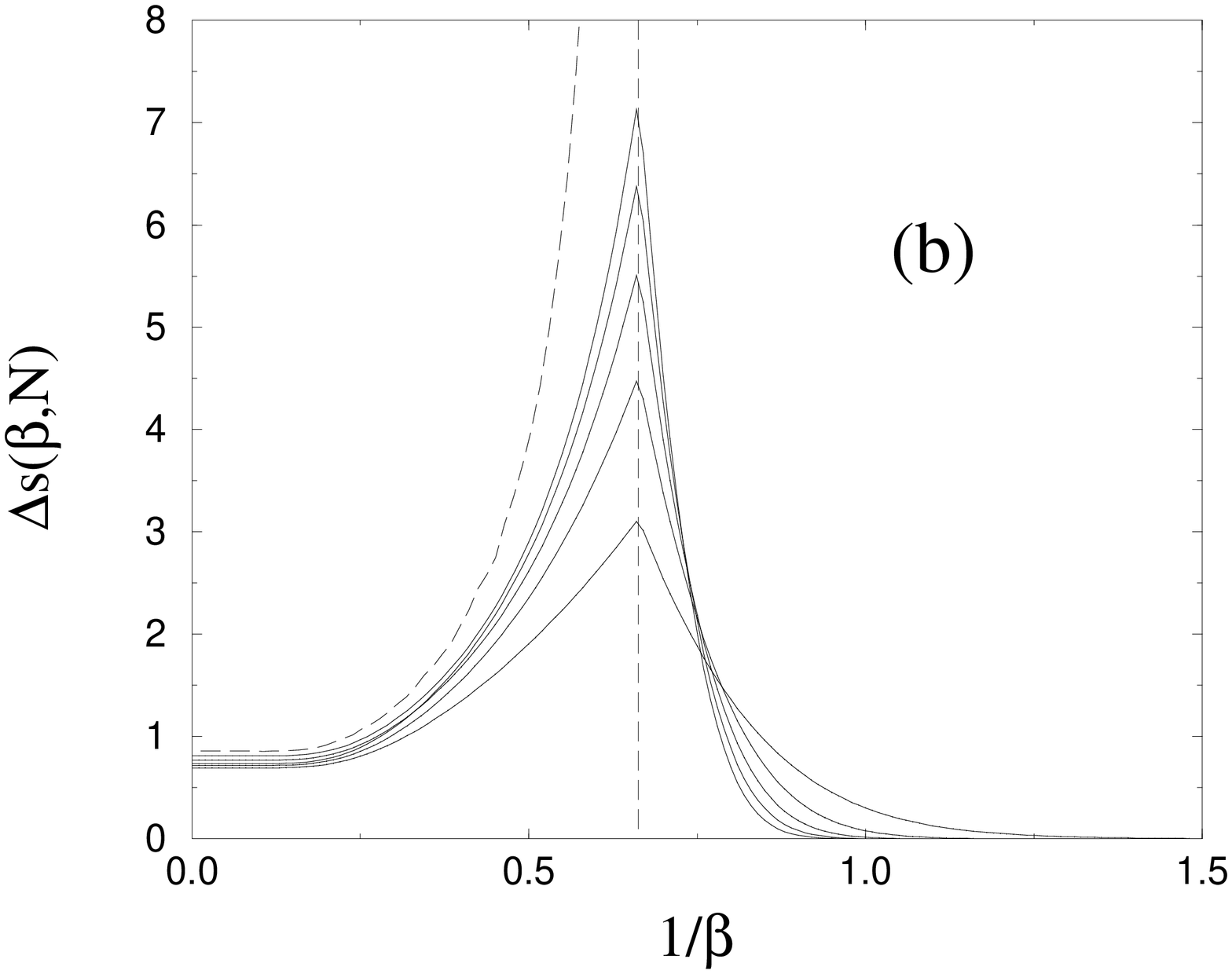,angle=-90,width=0.48\linewidth}
\end{tabular}
\caption{Finite size correction to the free energy (a) and to the
entropy (b) of the RCM. The continuous lines are the results of
numerical computations for $N=40,80,120,160,200$ (error bars are
not visible on this scale).
The dashed lines are the analytical results for the leading
finite size correction, for $N=40,200$ (a) and $N=200$ (b).}
\label{RCMFSS}
\end{figure}
In Fig. \ref{RCMFSS}, graph (a), we plot the quantity $\Delta f(\beta,N) \equiv
[f(\beta,N)-f_0(\beta)]N$, together with the theoretical 
prediction  $f_1(\beta,N)$ for several values of $N$.
In Fig. \ref{RCMFSS}, graph (b), we consider the entropy density 
$s(\beta,N) \equiv\beta^2\partial_{\beta} f(\beta,N)$:
we plot the difference  $\Delta s(\beta,N) \equiv
[s(\beta,N)-s_0(\beta)]N$, for the same values of $N$, 
together with  $s_1(\beta,N)\equiv\beta^2\partial_{\beta}f_1(\beta,N)$
for $N =200$ (the $N$ dependence of $s_1(\beta,N)$ is rather weak).

Two remarks can be made by looking at Fig, \ref{RCMFSS}. 
First, the $O(1/N^2)$ terms
in Eq. (\ref{FreeFS}) seems to be rather small. If the temperature is
not too close to the critical point, the finite size corrections are
well described by $f_1(\beta,N)$.
Second, the curves for $\Delta f(\beta,N)$, see Fig. 
\ref{RCMFSS}, graph (a), seem to cross 
at the critical point. This is expected since 
$\Delta f(\beta,N)\sim (1/2\beta_c)\log N$ for $\beta>\beta_c$, and
$\Delta f(\beta,N)\sim e^{-\kappa N}$  for $\beta<\beta_c$.
The crossing point $\beta_{N,N'}$ between the curves
$\Delta f(\beta,N)$ and $\Delta f(\beta,N')$ can be used to estimate $\beta_c$.
From the data of Fig. \ref{RCMFSS} 
we get 
\begin{eqnarray}
\beta_{40,80} = 1.52(1)\, ,\;\;
\beta_{80,120} = 1.51(1)\, ,\;\;
\beta_{120,160} = 1.51(1)\, ,\;\;
\beta_{160,200} = 1.51(1)\, ,
\end{eqnarray}
which is in good
agreement with the exact result $\beta_c \approx 1.50794$.

\subsection{The $(6,3)$ model}

In this case we are forced to consider quite small systems since
we do not know any simple form for the probability distribution of the
energy spectrum. We must enumerate all the codewords (i.e. the spin
configurations which satisfy the constraints in
Eq. (\ref{SpinwiseModel})): this takes at least $O(2^{NR})$ operations.
Notice that {\it finding} the codewords is a simple task. It suffices 
to solve the linear system ${\mathbb C}\ux = 0$  $({\rm mod} 2)$. A
standard method (we used gaussian elimination) takes $O(N^3)$
operations \cite{NR}.

\begin{figure}
\begin{tabular}{cc}
\hspace{-0.6cm}\epsfig{figure=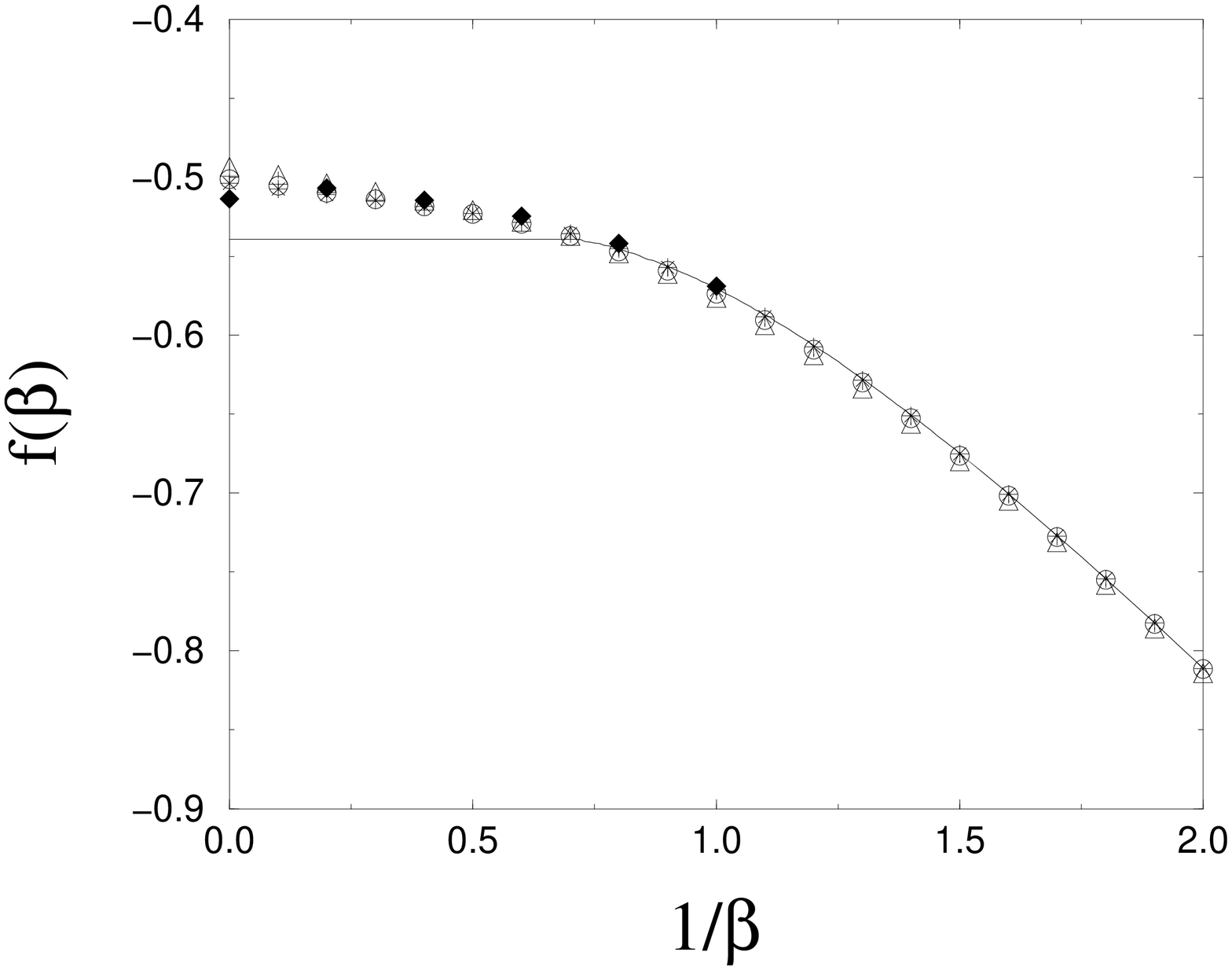,angle=-90,width=0.5\linewidth}&
\epsfig{figure=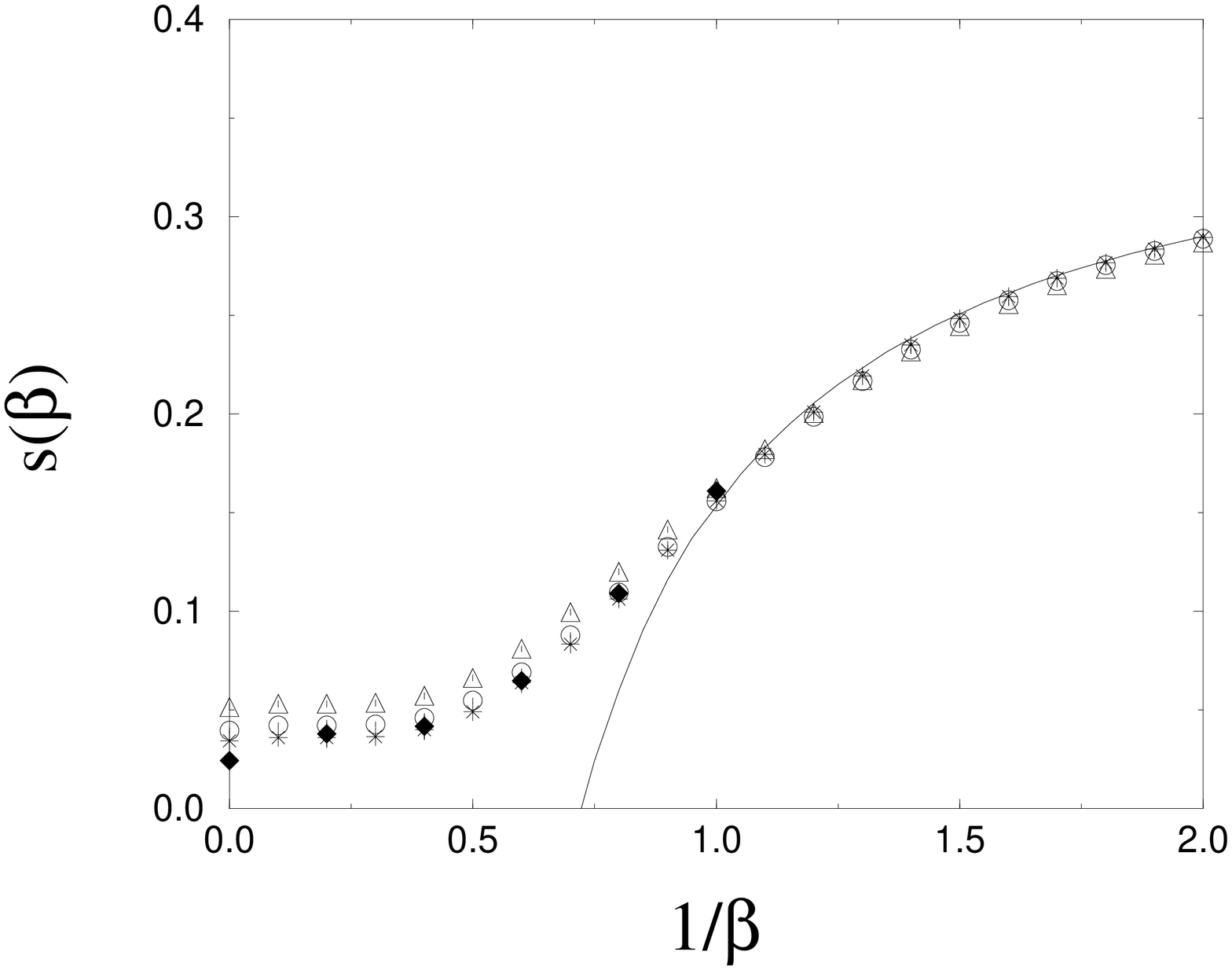,angle=-90,width=0.50\linewidth}
\end{tabular}
\caption{The free energy (left) and the entropy (right) of the $(6,3)$
model computed by exact-enumeration (symbols), and the corresponding theoretical
predictions (continuous lines). The various symbols refer to different 
system sizes: $N=20$ (triangles), $30$ (circles), $40$ (stars) and
$50$ (filled diamonds).}
\label{Enumeration}
\end{figure}
As in the previous Subsection, we fixed considered the binary 
field distribution (\ref{BinaryDistribution}) with $h_0 =
\atanh(1-2p)$, and $p = 0.2$. In Fig. \ref{Enumeration} we 
plot the results for the free energy and the entropy densities
for systems of size $N=20,30,40$  (averaged over $N_{stat} = 1000$
samples) and $N=50$ (with $N_{stat} = 20$ samples). 
The numerical results converge quite well to the theoretical 
calculation at high temperature. Below the critical temperature
the convergence is very slow, as expected from the analogy 
with the RCM example.

The sizes considered here are too small to reach any definite
conclusion on the glassy phase.
%
%
\section{Discussion}
\label{DiscussionSection}

The main result of this paper is the determination of the phase
diagram of regular Gallager codes, see Eq. (\ref{SpinwiseModel}).
This is depicted in Fig. \ref{BscGaussianPhases} for the infinite
connectivity limit. The phase diagram for finite connectivities has
been obtained by resorting to the replica method and looks
qualitatively similar. The most important quantitative difference 
is the critical noise level for the ferromagnetic-spin glass phase
transition. This quantity determines the performances of the corresponding code.
It can be determined either by solving the mean field equations
numerically, see Sec. \ref{ReplicaSection}, or in a large connectivity 
expansion, see Sec. \ref{ExpansionSection}. The result of the last
computation is reported in Fig. \ref{BscPhasesExp}.

The replica computation was made possible by the particularly simple
one-step replica symmetry breaking solution exhibited in
Eq. (\ref{RSBreaking}). We weren't able to prove that 
the saddle point (\ref{RSBreaking}) is either unique or the dominant
one.
There are however several independent indications
which confirm this conclusion:
\begin{itemize}
\item The proposed solution is consistent with the absence of 
replica symmetry breaking on the $\beta = 1$ line, which has been
proved in Sec. \ref{NishimoriSection}.
\item It has been shown \cite{SaadTighter,Theorem}
that the critical noise level is the same 
both for zero-temperature and for temperature one decoding.
This implies that the ferromagnetic-spin glass phase boundary must
pass through the points $(p = p_c(k,l), 1/\beta =0)$, and 
$(p = p_c(k,l), 1/\beta =1)$, see Fig. \ref{BscPhasesExp} (for sake of
simplicity we referred to the case of a binary field distribution).
This consistent with our phase diagram.
\item Our numerical results, although we restricted to fairly small
systems, do not contradict our conclusions.
\end{itemize}
It can be interesting to notice that recently \cite{FranzExact}
a ``factorized ansatz''
has been proposed as an exact one-step replica symmetry breaking solution
for some diluted spin models.
The solution used in this paper is, in some sense, complementary to the
one of Ref. \cite{FranzExact}.
%
%
\section*{Acknowledgments}

I am grateful to B.~Derrida for an illuminating discussion on the 
random codeword model, and to N.~Sourlas for his constant support and
encouragement.
I thank M.~M\'ezard and G.~Parisi for their interest in the subject of
this paper.
This work was supported through a European Community Marie Curie Fellowship.
%
%
\begin{appendix}

\section{Codewords in the $k,l\to \infty$ limit}
\label{CWProbAppendix}

In this Appendix we compute the one-codeword, and two-codeword
probabilities, see Eqs. (\ref{OneCodewordProbability}) and
(\ref{TwoCodewordProbability}), for generic values of $k$ and $l$.
Then we show that, in the $k,l\to \infty$ limit, different codewords
become statistically independent, i.e. $P_{\us,\ut}\sim P_{\us}P_{\ut}$.

The one-codeword probability is, to the leading exponential order:
\begin{eqnarray}
P_{\us} \sim \int\! \prod_{\sigma}d\l(\sigma)d\lh(\sigma)\,
\exp\{N A_1(\l,\lh;c)\}\, ,
\label{OneCodewordIntegral}
\end{eqnarray}
where
\begin{eqnarray}
A_1(\l,\lh;c) & =& -l\sum_{\sigma}\l(\sigma)\lh(\sigma)+
\frac{l}{2k}\left[\left(\sum_\sigma\l(\sigma)\right)^k+
\left(\sum_\sigma\l(\sigma)\sigma\right)^k
\right]+\nonumber\\
&&+l\sum_{\sigma}c(\sigma)\log\lh(\sigma)+l-\frac{l}{k}\, ,
\label{OneCodewordAction}
\end{eqnarray}
and $c(\sigma) = (1/N)\sum_i\delta_{\sigma,\sigma_i}$ characterizes
the configuration $\us$.
The above result can be proved by noticing 
that $\sum_{\us}P_{\us}\exp( \beta h_0\sum_i\sigma_i ) =
\<Z(h_0)\>_{\mathbb C}$, where $Z(h_0)$ is the partition function 
for the model (\ref{SpinwiseModel}) with uniform magnetic field $h_i =
h_0$. The average $\<Z(h_0)\>_{\mathbb C}$ is easily obtained from 
Eqs. (\ref{Zn}) and (\ref{ReplicatedAction}) by setting $n=1$ and
$p_h(h_i) = \delta(h_i-h_0)$.

The integral (\ref{OneCodewordIntegral}) can be done through the saddle
point method. Saddle point equations are more conveniently written
by eliminating $\lh(\us)$, and
using the variables $\l_+\equiv\sum_{\sigma}\l(\sigma)$
and  $\l_-\equiv\sum_{\sigma}\l(\sigma)\sigma$. We get:
\begin{eqnarray}
\l_+^k+\l_-^k & = &2\, ,\\
\l_-\l_+^{k-1}+\l_+\l_-^{k-1} &=& 2m\, ,
\end{eqnarray}
where $m = \sum_{\sigma}c(\sigma)\sigma=(1/N)\sum_i\sigma_i$.
For large  $k$, these equations imply $\l_+ = 2^{1/k}+O(m^k)$, $\l_- =
2^{1/k}m+O(m^k)$, as soon as $-1<m<1$. Substituting in 
Eq. (\ref{OneCodewordAction}), we get the result anticipated in
Sec. \ref{RandomCodewordSection}, see Eqs. (\ref{OCWresult1}),
(\ref{OCWresult2}).

Let us now consider the two-codeword probability, cf. Eq.
(\ref{TwoCodewordProbability}). Analogously to 
Eq. (\ref{OneCodewordIntegral}) we get:
\begin{eqnarray}
P_{\us,\ut} \sim \int\! \prod_{\sigma,\tau}
d\l(\sigma,\tau)d\lh(\sigma,\tau)\,
\exp\{N A_2(\l,\lh;c)\}\, .
\label{TwoCodewordIntegral}
\end{eqnarray}
The corresponding ``action'' is
\begin{eqnarray}
A_2(\l,\lh;c) & =& -l\sum_{\sigma,\tau}\l(\sigma,\tau)\lh(\sigma,\tau)+
\frac{l}{k}\sum_{\sigma_1\dots\sigma_k}\!\!^{'}
\sum_{\tau_1\dots\tau_k}\!\!^{'}
\l(\sigma_1,\tau_1)\dots\l(\sigma_k,\tau_k)+\nonumber\\
&&+l\sum_{\sigma,\tau}c(\sigma,\tau)\log\lh(\sigma,\tau)+l-\frac{l}{k}\, ,
\label{TwoCodewordAction}
\end{eqnarray}
where $c(\sigma,\tau)= (1/N)\sum_i\delta_{\sigma_i,\sigma}
\delta_{\tau_i,\tau}$, and the sums $\sum'$ are restricted
to $\sigma_1\cdots\sigma_k=+1$ and $\tau_1\cdots\tau_k=+1$.
As before we notice that 
$\sum_{\us,\ut}P_{\us,\ut}\exp( \beta h_1\sum_i\sigma_i+
\beta h_2\sum_i\tau_i )=\<Z(h_1)Z(h_2)\>_{\mathbb C}$ can be obtained
through a standard replica calculation, see Sec. \ref{ReplicaSection}
and App. \ref{ReplicaAppendix}, 
with $n = 2$ replicas. 

We now define the variables
$\l_0\equiv\sum_{\sigma,\tau}\l(\sigma,\tau)$,
$\l_{\sigma}\equiv\sum_{\sigma,\tau}\l(\sigma,\tau)\sigma$,
$\l_{\tau}\equiv\sum_{\sigma,\tau}\l(\sigma,\tau)\tau$,
and $\l_{\sigma\tau}\equiv\sum_{\sigma,\tau}\l(\sigma,\tau)\sigma\tau$.
The saddle point equations can be written in terms of these variables
as follows:
\begin{eqnarray}
\l_0^k+\l_{\sigma}^k+\l_{\tau}^k+\l_{\sigma\tau}^k & = & 4\, ,
\label{Z2Saddle1}\\
\l_{\sigma}\l_0^{k-1}+\l_0\l_{\sigma}^{k-1}
+\l_{\sigma\tau}\l_{\tau}^{k-1}+\l_{\tau}\l_{\sigma\tau}^{k-1} 
& = & 4m_{\sigma}\, ,\\
\l_{\tau}\l_0^{k-1}+\l_{\sigma\tau}\l_{\sigma}^{k-1}
+\l_0\l_{\tau}^{k-1}+\l_{\sigma}\l_{\sigma\tau}^{k-1} 
& = & 4m_{\tau}\, ,\\
\l_{\sigma\tau}\l_0^{k-1}+\l_{\tau}\l_{\sigma}^{k-1}
+\l_{\sigma}\l_{\tau}^{k-1}+\l_0\l_{\sigma\tau}^{k-1} 
& = & 4q\, ,\label{Z2Saddle4}
\end{eqnarray}
where $m_{\sigma} = \sum_{\sigma,\tau}c(\sigma,\tau)\sigma =
(1/N)\sum_i\sigma_i$,
$m_{\tau} = \sum_{\sigma,\tau}c(\sigma,\tau)\tau = (1/N)\sum_i\tau_i$,
and $q = \sum_{\sigma,\tau}c(\sigma,\tau)\sigma\tau = 
(1/N)\sum_i\sigma_i\tau_i$.
From Eqs. (\ref{Z2Saddle1})-(\ref{Z2Saddle4}),
we get, for $k\to\infty$, $\l_0\simeq 4^{1/k}$, 
$\l_{\sigma}\simeq 4^{(1-k)/k}m_{\sigma}$, 
$\l_{\tau}\simeq 4^{(1-k)/k}m_{\tau}$,
$\l_{\sigma\tau}\simeq 4^{(1-k)/k}q$, 
as soon as $-1<m_{\sigma},m_{\tau},q<1$.
The corrections to this asymptotic behavior are of 
order $O(m_{\sigma}^k,m_{\tau}^k,q^k)$.
Substituting this solution in Eqs. (\ref{TwoCodewordIntegral}),
(\ref{TwoCodewordAction}), we get the results (\ref{TCWresult1}),
(\ref{TCWresult2}).
%
%
\section{The random codeword model for a generic field distribution}
\label{RCMAppendix}

In this Appendix we 
solve\footnote{I am deeply indebted with B.~Derrida who explained to
me how to treat this general case.} the RCM for a generic field distribution 
$p_h(h_i)$.
The strategy is to start from a discrete distribution 
\begin{eqnarray}
p_h(h_i) = \sum_{q=1}^{\cal M} p_q\, \delta(h_i-h^{(q)})\, ,
\label{DiscreteFieldDistribution}
\end{eqnarray}
and then approximate a generic $p_h(h_i)$ by letting ${\cal M}\to\infty$.

Let us consider the distribution (\ref{DiscreteFieldDistribution}).
In the typical sample there will be $N_1\approx N p_1$ sites with
field $h_i=h^{(1)}$ (which we can suppose, without loss of generality, 
to be the sites $i=1,\dots,N_1$), $N_2\approx N p_2$ sites with
field $h_i=h^{(2)}$ (let us say for $i=N_1+1,\dots,N_1+N_2$), and so on.
For a given spin configuration $\us$, 
we define the partial magnetization $m_q(\us)$ as the magnetization
of the sites whose magnetic field is $h^{(q)}$. 
With the labeling of the sites chosen above we get
\begin{eqnarray}
m_q(\us) \equiv \frac{1}{N_q}\sum_{i = {\cal N}_{q-1}+1}^{{\cal N}_q}
\!\sigma_i\, ,
\end{eqnarray}
where ${\cal N}_q = N_1+\dots+N_q$. 
We call $\{m_q(\us)\}$ the {\it magnetization profile} of the
configuration $\us$.

\begin{figure}
\centerline{
\hspace{-1.5cm}\epsfig{figure=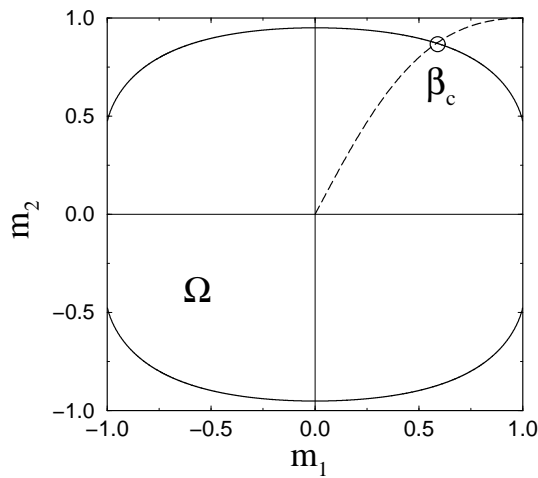,angle=-90,width=0.4\linewidth}}
\caption{The RCM for 
$p_h(h_i) = (2/5)\, \delta(h_i-1/2)+(3/5)\, \delta(h_i-1)$.
The continuous line encircles the region $\Omega$ (see text).
The dashed line is the curve $m_1 = \tanh\beta/2$, $m_2 = \tanh\beta$,
which intersect the boundary of $\Omega$ for $\beta = \beta_c$.}
\label{RCMFigure}
\end{figure}
We now consider the $2^{NR}$ states $\alpha = 1,\dots, 2^{NR}$.
To each of them it is associated a random codeword $\us^{(\alpha)}$,
where the $\sigma_i^{(\alpha)}$ are quenched variables drawn with flat
probability distribution.
We ask ourselves what is the typical number 
${\cal N}_{typ}(\{m_q\})$ of states $\alpha$ having a
given magnetization profile $m_q(\us^{(\alpha)})=m_q$. 
The answer is quite easy.
Define the function ${\cal G}(\{m_q\})$ as follows
\begin{eqnarray}
{\cal G}(\{m_q\}) = R\log 2+\sum_{q=1}^M p_q {\cal H}(m_q)\, ,
\label{TypicalNumber}
\end{eqnarray}
where ${\cal H}(x)$ is given in Eq. (\ref{EntropyDefinition}).
The typical number ${\cal N}_{typ}(\{m_q\})$ is obtained from 
${\cal G}(\{m_q\})$ through the usual construction: 
${\cal N}_{typ}(\{m_q\})\sim \exp[ N{\cal G}(\{m_q\})]$
if ${\cal G}(\{m_q\})>0$ and ${\cal N}_{typ}(\{m_q\}) = 0$ otherwise.
The convex region $\Omega\equiv\{\{m_q\}|{\cal G}(\{m_q\})>0\}$ 
is depicted in Fig. \ref{RCMFigure} for the
case ${\cal M}=2$.

The energy of a state $\alpha$ can be written in terms of
its magnetization profile: $E^{(\alpha)} = -N\sum_q p_q
h^{(q)}m_q(\us^{(\alpha)})$.
The free energy density can therefore computed from
${\cal N}_{typ}(\{m_q\})$ as follows:
\begin{eqnarray}
f(\beta) = \min_{\{m_q\} }\left\{
-\frac{1}{\beta}{\widehat {\cal G}}(\{m_q\})-\sum_{q=1}^M p_q h_q 
m_q\right\}\, ,
\label{FreeLegendre}
\end{eqnarray}
where ${\widehat {\cal G}}(\{m_q\})\equiv (1/N)\log {\cal
N}_{typ}(\{m_q\})$ (i.e. ${\widehat {\cal G}}(\{m_q\})={\cal
G}(\{m_q\})$ inside $\Omega$, and  ${\widehat {\cal
G}}(\{m_q\})=-\infty$ outside). 

If the expression (\ref{TypicalNumber}) is used in
Eq. (\ref{FreeLegendre}), one gets the saddle point condition 
$m_q = \tanh \beta h_q$. 
This describes a curve in the $\{m_q\}$ space which start at $m_q=0$
for $\beta =0$, and ends at $m_q = {\rm sign}\, h_q$ for $\beta = \infty$.
The corresponding free energy reads 
\begin{eqnarray}
f_P(\beta) = -\frac{R}{\beta}\log 2-\frac{1}{\beta}
\sum_{q=1}^M p_q \log\cosh \beta h_q\, .
\label{DiscreteFreeEnergy}
\end{eqnarray}
At some critical temperature $\beta = \beta_c$ the curve $m_q
=\tanh\beta h_q$ crosses the boundary of $\Omega$. The saddle point
$m_q=\tanh\beta h_q$ is no longer valid for $\beta>\beta_c$.
The critical temperature can be computed from the zero entropy
condition $\partial_{\beta}f_P|_{\beta = \beta_c} = 0$.
For $\beta>\beta_c$ the entropy vanishes and the free energy is
frozen to its value at the critical point:
$f_{SG}(\beta)=f_P(\beta_c)$.
As in Sec. \ref{RandomCodewordSection}, we must include in our analysis
the ordered state $\alpha = 0$ whose free energy is
$f_F(\beta) = -\<h\>_h$. 

The solution for a continuous field distribution $p_h(h_i)$ follows
from the above results by taking the ${\cal M}\to \infty$ limit in Eq.
(\ref{DiscreteFreeEnergy}). This yields Eq. (\ref{FreeRCWGeneric}).
Alternatively we could have started with a continuous magnetization
profile $m(h)$ from the very beginning of this Appendix.  
%
%
\section{The derivation of Eq. (\ref{ReplicatedAction})}
\label{ReplicaAppendix}

We start by writing down the partition function
of the model (\ref{SpinwiseModel}):
\begin{eqnarray}
Z(\beta) = \sum_{\us}\, \prod_{j=1}^M
\delta[\sigma^{\omega_j},+1]\, e^{\sum_i
h_i\sigma_i}\, .
\end{eqnarray}
We rewrite
the constraint term (i.e. the product of Kronecker delta functions)
by introducing the quenched variables $D_{\omega} = 0,1$,
where $\omega = (i^\omega_1,\dots,^\omega_k)$ runs over the
$k$-plets of site indices. 
The variables $D_{\omega}$ are defined by setting $D_{\omega} = 1$ if 
$\omega = \omega_j$ for some $j=1,\dots,M$ and $D_{\omega} = 0$
otherwise.
With this definition we can write the replicated partition function as
follows
\begin{eqnarray}
\<Z^n\> = \frac{1}{\cal N}\sum_{\{ D\}}\sum_{\{\ps\}}
\prod_{i=1}^N\left\<e^{\beta h\sum_a\sigma^a_i}\right\>_h
\prod_{\omega}\{1-D_{\omega}+D_{\omega}\delta_n[\ps^{\omega}]\}\, ,
\end{eqnarray}
where $\ps^{\omega} \equiv (\prod_{r=1}^k\sigma^1_{i^{\omega}_r},
\dots, \prod_{r=1}^k\sigma^n_{i^{\omega}_r})$, 
$\delta_n[\ps]\equiv \prod_{a=1}^n\delta[\sigma^a,+1]$,
and ${\cal N}$ is a normalization constant (to be computed later). 

According to our choice of the {\it ensemble} of check matrices,
we must impose $\sum_{\omega\ni i}D_{\omega} =l$, for any 
$i=1,\dots,N$.  This can be done by using the identity
\begin{eqnarray}
\delta\left[\sum_{\omega\ni i}D_{\omega},l\right] =
\oint\!\frac{dz_i}{2\pi i}\frac{1}{z_i^{l+1}}\, z_i^{\sum_{\omega\ni
i}D_{\omega}}\, ,
\end{eqnarray}
where the integration path encircles the origin in the complex $z_i$
plane. We get
\begin{eqnarray}
\<Z^n\> = \frac{1}{{\cal N}'}\sum_{\{\ps\}}
\prod_{i=1}^N \oint\!\frac{dz_i}{2\pi i}\frac{1}{z_i^{l+1}}\,
\left\<e^{\beta h\sum_a\sigma^a_i}\right\>_h
\prod_{\omega}\sum_{D_{\omega}=0}^1 w(D_{\omega})
\{1-D_{\omega}+D_{\omega}\delta_n[\ps^{\omega}]\}\, 
z_{\omega}^{D_{\omega}}\, ,\!\!\!\!\!\!\!\!\!\!\!\!\!\!\!\!\nonumber\\
\end{eqnarray}
where $z_{\omega}\equiv \prod_{i\in\omega}z_i$.
The weights $w(D_{\omega})$ have been introduced for later
convenience, and correspond to a rescaling of the $\{ z_i\}$. 
Their contribution can be readsorbed by the 
normalization constant ${\cal N}'$.
We set $w(1) = l(k-1)!/N^{k-1}$ and $w(0) = 1-w(1)$.
Now we can sum over the $D_{\omega}$, obtaining
\begin{eqnarray}
\<Z^n\> & =& \frac{1}{{\cal N}''}\sum_{\{\ps\}}
\prod_{i=1}^N \oint\!\frac{dz_i}{2\pi i}\frac{1}{z_i^{l+1}}\,
\left\<e^{\beta h\sum_a\sigma^a_i}\right\>_h\cdot\\
&&\phantom{\frac{1}{{\cal N}''}\sum_{\{\ps\}}
\prod_{i=1}^N \oint\!\frac{dz_i}{2\pi i}}
\cdot\exp\left\{\frac{Nl}{k}\sum_{\ps_1,\dots,\ps_k}
c_z(\ps_1)\dots c_z(\ps_k)
\prod_{a=1}^n \delta[\sigma^a_1\dots\sigma^a_k,+1]\right\}\, ,
\nonumber
\end{eqnarray}
where $c_z(\ps)\equiv(1/N)\sum_i z_i \delta_{\ps,\ps_i}$.
Finally we introduce the order parameter $\l(\ps)$ and 
its complex conjugate $\lh(\ps)$, by using the following identity
\begin{eqnarray}
\exp\{N{\cal F}[c]\} &=& 
\int\!\prod_{\ps}\frac{Nl}{\pi} d\l(\ps)d\lh(\ps)\,
\exp\left\{-Nl\sum_{\ps}\l(\ps)\lh(\ps)+\right.\\
&&\phantom{\int\!\prod_{\ps}\frac{Nl}{\pi} d\l(\ps)d\lh(\ps)
\exp\left\{\right.}\left.+N{\cal F}[\l]+
Nl\sum_{\ps}\lh(\ps)c_z(\ps)\right\}\, .\nonumber
\end{eqnarray}
The use of the above identity allows to integrate over the $\{z_i\}$,
obtaining Eqs. (\ref{Zn}) and (\ref{ReplicatedAction}).
The overall normalization constant can be fixed by 
requiring $\<Z^n\>\sim 2^{Nn(1-l/k)}$ for $h_i=0$.
%
%
\section{Large $k,l$ expansion: general formulae}
\label{FormulaeAppendix}

Let us define $t_p\equiv \<\tanh \beta h\>_h$. We assume formally
$t_p = O(t^p)$ where $t$ is ``small'' and expand in $t^k$ to the order
$t^{3k}$. All the observables can be expressed in terms of the order
parameters $\pi(x)$ and $\ph(y)$. The solutions of 
Eqs. (\ref{SaddleRS1}), (\ref{SaddleRS2}) admit an expansion of the form
\begin{eqnarray}
\pi(x) = p_h(x)+\sum_{m=1}^{\infty} \pi_m \beta^{-m} p_h^{(m)}(x)\;\;
;\;\;\;\;\; \ph(y) = \delta(y)+\sum_{n=1}^{\infty}\ph_n \beta^{-n}
\delta^{(n)}(y)\, , 
\end{eqnarray}
where $p_h^{(m)}(x) \equiv \partial_x^m p_h(x)$ and 
$\delta^{(n)}(y) = \partial_y^n \delta(y)$. Moreover one gets 
$\pi_m,\ph_m = O(t^{mk})$. The results for the first few coefficients
are listed below:
{\footnotesize
\begin{eqnarray}
\pi_1 & = & -(l-1)t_1^{k-1}-(k-1)(l-1)^2(1-t_2)t_1^{2k-3}- \\
&&- \frac{1}{3}(l-1)t_3^{k-1} - \frac{1}{2}(k-1)(k-2)(l-1)^3(1-t_2)^2t_1^{3k-5}
-(k-1)^2(l-1)^3(1-t_2)^2t_1^{3k-5} +\nonumber\\
&&+ (k-1)(l-1)^2(t_1-t_3)t_2^{k-1}t_1^{k-2}
+ (k-1)(l-1)^2(l-2)(t_1-t_3)t_1^{3k-4} + O(t^{4k})\, ,\nonumber\\
\pi_2 & = & \frac{1}{2}(l-1)t_2^{k-1}+\frac{1}{2}(l-1)(l-2)t_1^{2k-2}+\\
&&+(k-1)(l-1)^2(t_1-t_3)t_2^{k-2}t_1^{k-1}+
(k-1)(l-1)^2(l-2)(1-t_2)t_1^{3k-4} + O(t^{4k})\, ,\nonumber\\
\pi_3 & = &
-\frac{1}{6}(l-1)t_3^{k-1}-\frac{1}{2}(l-1)(l-2)t_2^{k-1}t_1^{k-1}-
\frac{1}{6}(l-1)(l-2)(l-3)t_1^{3k-3}+O(t^{4k})\, ,\\
\nonumber\\
\ph_1 & = & -t_1^{k-1} - (k-1)(l-1)(1-t_2)t_1^{2k-3}-\\
&& -\frac{1}{2}(k-1)(k-2)(l-1)^2(1-t_2)^2t_1^{3k-5} -
(k-1)^2(l-1)^2(1-t_2)^2t_1^{3k-5}+ \nonumber\\
&&+(k-1)(l-1)(t_1-t_3)t_2^{k-1}t_1^{k-2}+
(k-1)(l-1)(l-2)(t_!-t_3)t_1^{3k-4}-\frac{1}{3}t_3^{k-1}+O(t^{4k})\, ,
\nonumber\\
\ph_2 & = &
\frac{1}{2}t_2^{k-1}+(k-1)(l-1)(t_1-t_3)t_2^{k-2}t_1^{k-1}+O(t^{4k})\\
\ph_3 & = & -\frac{1}{6}t_3^{k-1}+O(t^{4k})\, .
\end{eqnarray}
}

The result for the paramagnetic free energy is
{\footnotesize
\begin{eqnarray}
\beta f_P(\beta) & = & -R\log 2-\<\log\cosh \beta h\>_h
-\frac{l}{k}t_1^k-\frac{1}{2}l(l-1)(1-t_2)t_1^{2k-2}
+\frac{1}{2}\frac{l}{k}t_2^k-\nonumber\\
&& -\frac{1}{2}(k-1)l(l-1)^2(1-t_2)^2t_1^{3k-4}+
\frac{1}{3}l(l-1)(l-2)(t_1-t_3)t_1^{3k-3}+
\label{FreeExpansion}\\
&&+l(l-1)(t_1-t_3)t_1^{k-1}t_2^{k-1}-\frac{1}{3}\frac{l}{k}t_3^k
+O(t^{4k})\, .
\nonumber
\end{eqnarray}
}
%
%
\section{Finite size corrections for the random codeword model}
\label{FiniteSizeAppendix}

Let us consider the binary field distribution
(\ref{BinaryDistribution}) with $h_0 =1$. The results for a generic
value of $h_0$ are obtained after a trivial rescaling of
energies and temperatures: $f(\beta,h_0;N) = h_0 f(\beta h_0,1; N)$.
 
As explained in Sec. \ref{FiniteSizeSection}, the finite size
corrections at the paramagnetic-spin glass phase transition can be
studied by neglecting the ordered state. This introduces exponentially
small errors. 
The calculation of the free energy can be done along the lines of
Ref. \cite{DerridaREM}, Appendix B, which starts from the identity:
\begin{eqnarray}
\<\log Z\> = \int_0^{\infty}\frac{dt}{t}\left(e^{-t}-e^{-tZ}\right)\, .
\end{eqnarray}

We limit ourselves to quoting the outcome of the calculation.
For $\beta< \beta_c$, we get $f(\beta,N)= f_P(\beta)+O(e^{-\kappa
N})$\footnote{Obviously the ordered state cannot be longer neglected in
computing $\kappa$}. For $\beta >\beta_c$ we get Eq. (\ref{FreeFS}),
with 
\begin{eqnarray}
f_0(\beta) = -\widehat{\epsilon}(R)\, ,\;\;\;
f_1(\beta,N) = \int_0^{\infty} \! d\phi\, \rho(\phi) \, e^{-\phi} +
\gamma/\beta\, ,
\end{eqnarray}
$\gamma \approx 0.577216$ being the Euler constant.
The function $\rho(\phi)$ is defined as the (unique) solution of
\begin{eqnarray}
\beta_c \rho + \log \Psi(-N\widehat{\epsilon}+\rho) = \log(\phi)
+\frac{1}{2}\log\left[\frac{\pi}{2}N(1-\widehat{\epsilon}^2)\right]\, ,
\label{RhoEq}
\end{eqnarray}
where $-\widehat{\epsilon}(R)$ is the ground state energy density in the
thermodynamic limit, see Sec. \ref{RandomCodewordSection}. 
The function $\Psi(x)$ is defined as follows
\begin{eqnarray}
\Psi(x) = \sum_{q = -\infty}^{+\infty}e^{-\beta_c(2q+x)}
\left[1-\exp\left(-e^{\beta(2q+x)}\right)\right]\, .
\label{PsiDefinition}
\end{eqnarray}
Notice that $\Psi(x+2) = \Psi(x)$. The $\log \Psi$ term in 
Eq. (\ref{RhoEq}) gives therefore an oscillating $N$ dependence to 
$f_1(\beta,N)$. Moreover, since $\Psi(-N\widehat{\epsilon}+\rho)$
remains finite 
for any $N$ and $\rho$, $f_1(\beta,N) \sim (1/2\beta_c)\log N$ as
$N\to\infty$.
Finally we remark that the sum in Eq. (\ref{PsiDefinition}) diverges
as $\beta\downarrow \beta_c$. This gives the singularity of the free
energy corrections at the critical point:
$f_1(\beta,N) \sim (1/\beta_c)\log(1-\beta_c/\beta)$.

\end{appendix}


\begin{thebibliography}{99}

\bibitem{Cover} T.~M.~Cover and J.~A.~Thomas, 
{\it Elements of Information Theory}, (Wiley, New York, 
1991).

\bibitem{Viterbi} A.~J.~Viterbi and J.~K.~Omura,
{\it Principles of Digital Communication and Coding},
(McGraw-Hill, New York, 1979).

\bibitem{Shannon} C.~E.~Shannon, Bell Syst. Tech. J. {\bf 27}, 
379-423, 623-656  (1948).

\bibitem{Chung} S.-Y.~Chung, G.~D.~Forney,~Jr., T.~J.~Richardson and 
R.~Urbanke, 
{\it On the design of low-density parity-check codes within $0.0045$ dB from
the Shannon limit}, IEEE Comm. Letters, to appear. 

\bibitem{PrimoBerrou} C.~Berrou, A.~Glavieux, and
P.~Thitimajshima. Proc. 1993 Int. Conf. Comm. 1064-1070.

\bibitem{MacKay} D.~J.~.C.~MacKay, IEEE Trans. Inform. Theory {\bf 45},
399-431 (1999).

\bibitem{Gallager} R.~G.~Gallager. {\it Low Density Parity Check Codes},
Research Monograph Series Vol. 21 (MIT, Cambridge, MA., 1963).

\bibitem{Sourlas1} N.~Sourlas. Nature {\bf 339}, 693-694 (1989).

\bibitem{Sourlas2} N.~Sourlas, 
{\it Statistical Mechanics of Neural Networks} 
Lecture Notes in Physics {\bf 368}, edited by L.~Garrido 
(Springer Verlag, 1990).

\bibitem{Sourlas4} N.~Sourlas, {\it From Statistical Physics to 
 Statistical Inference and Back,} edited by 
P.~Grassberger and J.-P.~Nadal (Kluwer Academic, 1994), p. 195.

\bibitem{Turbo1} A.~Montanari and N.~Sourlas, Eur.~Phys.~J.~B {\bf
18}, 107-119 (2000).

\bibitem{Turbo2} A.~Montanari, Eur.~Phys.~J.~B {\bf 18}, 121-136 (2000).

\bibitem{KanterSaad_PRL} I.~Kanter and D.~Saad, Phys.~Rev.~Lett. {\bf
83}, 2660-2663 (1999).

\bibitem{KanterSaad_Cascading} I.~Kanter and D.~Saad, 
Phys.~Rev.~E. {\bf 61}, 2137-2140 (1999). 

\bibitem{Saad_MN_PRL} Y.~Kabashima, T.~Murayama and D.~Saad,
Phys.~Rev.~Lett. {\bf 84}, 1355-1358 (2000).

\bibitem{KanterSaad_FS} I.~Kanter and D.~Saad,
Jour.~Phys.~A. {\bf 33}, 1675-1681 (2000).

\bibitem{SaadRegular} R.~Vicente, D.~Saad and Y.~Kabashima,
Phys.~Rev.~E. {\bf 60}, 5352-5366 (1999). 

\bibitem{SaadCactus} R.~Vicente, D.~Saad and Y.~Kabashima.
Europhys. Lett. {\bf 51}, 698-704 (2000?). 

\bibitem{SaadTighter} Y.~Kabashima, N.~Sazuka, K.~Nakamura and
D.~Saad, {\it Tighter Decoding Reliability Bound for Gallager's 
Error-Correcting Code}, {\tt cond-mat/0010173}.

\bibitem{Nishimori1} H.~Nishimori. J.~Phys.~C {\bf 13}, 4071-4076
(1980).

\bibitem{DerridaREM} B.~Derrida. Phys.~Rev.~B {\bf 24}, 2613-2626 (1981).

\bibitem{SpinGlass} M.~Mezard, G.~Parisi and M.~A.~Virasoro, {\it Spin
Glass theory and Beyond.} (World Scientific, Singapore, 1987).

\bibitem{MonassonRSB} R.~Monasson, J.~Phys.~A {\bf 31} (1998) 513-529.

\bibitem{Tanner} R.~M.~Tanner, IEEE~Trans.~Infor.~Theory, {\bf 27},
533-547 (1981).

\bibitem{Nishimori2} H.~Nishimori. Prog.~Theor.~Phys. {\bf 66}, 
1169-1181 (1981).

\bibitem{Nishimori3} H.~Nishimori and D.~Sherrington, 
{\it Absence of Replica Symmetry Breaking in a Region of the Phase
Diagram of the Ising Spin Glass}, {\tt cond-mat/0008139}.

\bibitem{Rujan} P.~Ruj{\'a}n, Phys.Rev.Lett. {\bf 70}, 2968-2971 (1993).

\bibitem{Sourlas5} N.~Sourlas. Europhys.Lett. {\bf 25}, 159-164 (1994).

\bibitem{RichardsonUrbanke} T.~Richardson and R.~Urbanke, {\it 
The Capacity of Low-Density Parity Check Codes under Message-Passing
Decoding}, IEEE~Trans.~Inform.~Theory, to appear.

\bibitem{Wong} K.~Y.~M.~Wong and D.~Sherrington, J.~Phys.~A {\bf 21}
L459-L466 (1988).

\bibitem{MezardParisiBethe} M.~M\'ezard and G.~Parisi, 
{\it The Bethe lattice spin glass revisited}, {\tt
cond-mat/0009418}, to appear in Eur.~Phys.~J.~B.

\bibitem{BiroliVariational} G.~Biroli, R.~Monasson, M.~Weigt, 
Eur.~Phys.~J.~B {\bf 14}, 551-568 (2000).

\bibitem{NR} W.~H.~Press, B.~P.~Flannery, S.~A.~Teukolsky and
W.~T.~Vetterling, {\it Numerical Recipes}, (Cambridge University Press, 
Cambridge, 1986).

\bibitem{Theorem} D.~J.~.C.~MacKay, {\it On thresholds of codes},
available at {\tt http://}{\tt wol.ra.phy.cam.ac.uk/}{\tt mackay/abstracts/theorems}. 

\bibitem{FranzExact} S.~Franz, M.~Leone, F.~Ricci-Tersenghi and
R.~Zecchina, {\it Exact solutions for diluted spin glasses and
optimization problems}, {\tt cond-mar/0103328}. 

\end{thebibliography}
\end{document}